\documentclass{aa}
\usepackage{graphicx,natbib}
\newcommand {\BS}{{{\it Beppo}SAX}\ \ignorespaces}
\newcommand {\xmm}{{XMM-\it Newton}\ \ignorespaces}
\def\gta{ \lower .75ex \hbox{$\sim$} \llap{\raise .27ex \hbox{$>$}} }
\def\lta{ \lower .75ex\hbox{$\sim$} \llap{\raise .27ex \hbox{$<$}} }
\def\r{{\sc REF\ \ignorespaces}}
\def\a{{\sc ABS\ \ignorespaces}}
\begin{document}

\subtitle{}



\author{P.O.\,Petrucci\inst{1} \and G. Ponti\inst{2,3} \and
  G.\,Matt\inst{4} \and A.L.\,Longinotti\inst{5} \and J.\,Malzac\inst{6} \and M.\,Mouchet\inst{7,8} \and
C.\,Boisson\inst{7} \and 
L.\,Maraschi\inst{9} \and K.\,Nandra\inst{10} 
\and P.\,Ferrando\inst{8,11} }


\institute{Laboratoire d'Astrophysique de Grenoble, BP 43, 38041 Grenoble
  Cedex 9, France \and Dipartimento de Astronomia, Universit\`a degli Studi di Bologna, via Zamboni, I-40127 Bologna, Italy \and INAF-IASF Sezione di Bologna, Via Gobetti 101,
  I-40129 Bologna, Italy \and Dipartimento di Fisica, 
  Universit\`a degli Studi ``Roma tre'', via della Vasca Navale 84,
  I-00046 Roma, Italy \and XMM-Newton Science
  Operations Center, European Space Astronomy Center, ESA, 28080
  Madrid, Spain \and Centre d'\'etude Spatiale des Rayonnements
  (CNRS/UPS/OMP), 31028 Toulouse, France  \and LUTH, Observatoire de
  Paris, Section de 
  Meudon, 92195 Meudon Cedex, France \and APC Universite Paris 7 Denis Diderot F-75205 Paris Cedex 13   \and Osservatorio Astronomico di Brera, Via Brera 28,
  02121 Milano, Italy \and Astrophysics 
  Group, Imperial College London, Blackett Laboratory, Prince Consort
  Road, London SW7 2AW \and Service d'Astrophysique, DSM/DAPNIA/SAp,
  CE Saclay, 91191 Gif-sur-Yvette Cedex, France   }

\date{}

\title{Unveiling the broad band X-ray continuum and iron line complex in
 \object{Mkr 841}}

\abstract
{ \object{Mkr 841} is a bright Seyfert 1 galaxy known to harbor a strong soft
  excess and a variable K$\alpha$ iron line. }  
{ It has been observed during 3 different periods (January 2001,
  January 2005 and July 2005) by XMM for a total cumulated
  exposure time of $\sim$108 ks. We present in this paper a broad band
  spectral analysis of the complete EPIC-pn data
  sets.} 
{We use different methods of data analysis including model-independent methods (spectral ratios, RMS, ...) as well as model fitting. We were able to test two different models for the soft excess, a relativistically  blurred photoionized
  reflection (\r model) and a relativistically smeared ionized absorption (\a model). The continuum is modeled by a simple cut-off power law and we also add a neutral reflection.}
{These observations confirm the presence of a soft excess and iron line and
  reveal extreme and puzzling spectral and temporal
  behaviors. The 0.5-3 keV soft X-ray flux decreases by a factor
  3 between 2001 and 2005 and the line shape appears to be a mixture of
  broad and narrow components, the former being variable on small (ks) time scale while the later is consistent with being constant. The 2-10 keV spectrum also hardens between 2001 and 2005.  We succeed in describing this complex broad-band 0.5-10 keV spectral variability using  either \r or \a to fit the soft excess. Both models give statistically equivalent results  even including simultaneous \BS data up to 200 keV. Both models are consistent with the presence of remote reflection characterized by a constant narrow component in the data. However they differ in the presence of a broad line component  present in \r but not needed in {\sc ABS}. Consequently the physical interpretation of the line profile variability is quite different, resulting from the variability of the broad line component in \r and from the variability of the absorbing medium in {\sc ABS}.
 
%

This study also reveals the sporadic presence of relativistically redshifted narrow iron lines, one of them being detected at 4.8 keV in the EPIC-pn instruments at more than 98.5\% confidence level.  If interpreted as the blue horn of a relativistically distorted neutral iron line, the large redshift implies the presence of a Kerr black hole.}
{}
\maketitle

\begin{table*}
\begin{center}
\begin{tabular}{lccccc}
\hline
 & OBS 1 & OBS 2 & OBS 3 & OBS4 & OBS5 \\
\hline
Start date & 2001-01-13 & 2001-01-13 & 2001-01-14 & 2005-01-16 & 2005-07-17 \\
 & (05h20m55s UT) & (09h33m50s UT) & (00h52m28s UT) & (12h38m21s UT) & (06h38m03s UT)\\
Obs. ID & 0112910201 & 0070740101 & 0070740301 & 0205340201 & 0205340401\\
Total duration (s) & 10106 & 12336 & 14775 & 49500 & 29509\\
Good exp. time (\%) & 58 & 62 & 74 & 61 & 44\\ 
Cts.s$^{-1}$  & 17.9 & 22.2 & 21.8 & 5.6 & 7.2 \\
\hline
\end{tabular}
\caption{Observation epochs, total duration, \% of good exposure time and mean count rates in the EPIC-pn  instrument.\label{log}}
\end{center}

\end{table*}
\section{Introduction}
Seyfert 1 galaxies emit the bulk of their luminosity in the UV and X-ray
bands. Their spectral energy distributions are characterized by two main
components: a UV bump peaking in the UV range \citep{obr88,kin91}, and an apparently non-thermal X-ray power law
component extending up to at least 100 keV. A majority of objects show
also the presence of a soft X-ray excess with respect to the high energy
power law extrapolation (e.g. \citealt{wal93,pag04} ).\\

The origin of these components is not well understood. The basic paradigm
supposes the existence of an accreting supermassive black hole. The
gravitational energy released by the accreting gas is generally thought
to be dissipated partly in the UV as thermal heating in an optically thick ``cold'' plasma \citep{shi78,malsar82} and partly in
X-rays in active blobs of optically thin ``hot'' plasma (e.g. \citealt{haa94}). These blobs
radiate mainly through comptonization of soft photons, possibly provided
by the optically thick medium. Irradiation of the dense gas by hard
X--rays produces also an X--ray reflection spectrum. The study of this
reflected component (dominated by the iron K$\alpha$ emission line and
the reflection bump peaking between 20-40 keV) appears to be of great
importance since it has the potential to be a key diagnostic of the
strong gravity environment of black holes (Fabian et al. 2000 and
references therein).\\ On the other hand, while the origin of  the soft excess is
still not clearly understood, it was recently realized that its characteristic temperature (when fitted by e.g. a simple black body) is remarkably constant over a wide range of AGN luminosities and black hole masses (e.g. \citealt{cze03,gie04,cru06,pon06}), favoring an origin through atomic processes instead of  purely continuum emission.
For example, recent studies suggest that an appealing
explanation could be (photoionized) reflection from the accretion
disc \citep{cru06}. This spectral decomposition has been
successful in fitting the XMM-{\it Newton} X--ray data of many sources like MCG--6-30-15, 1H~0707--495, NGC~4051 or MCG-02-14-009 \citep{fab04,pon06,lar07,por06}. Moreover, in
those cases in which a broad Fe line is clearly detected (such as in
MCG--6-30-15) the model is very attractive because the soft excess and broad
Fe line are fitted self--consistently with the same relativistically
blurred reflection model. But absorption instead of reflection could also reproduce the soft excess \citep{gie04,gie06,sob07,sch06} and modify the spectral shape close to the iron line to mimic the presence of a broad component. Nevertheless the reality is almost certainly a complex combination of absorption and reflection effects (e.g. \citealt{che06}) and their relative importance in the observed spectra has been a significant topic of discussion in the recent literature. One of the main issues of this debate is the  determination of the underlying continuum below the iron line in order to permit a precise measurement of the line broadness, the most important signature of the presence of black holes in AGNs.\\

%
%
%

 \object{Mkr 841} is a bright Seyfert 1 galaxy (z=0.0365, \citealt{fal99}), one of the rare Seyfert
1s detected by OSSE at more than 3 $\sigma$ \citep{joh97,zdz00}. It is known
for its large spectral variability \citep{geo93,nan95}, its strong soft excess (this was the first object where a soft
excess was observed, \citealt{arn85}) and its variable iron line (at
least on a year time scale, \citealt{geo93}). The later was observed
in some cases with a relatively large equivalent width (hereafter EW) of
about 400 eV \citep{day90,bia01} significantly above the value predicted by standard cold reflection
model (e.g. \citealt{geo91}).\\

Recent XMM-{\it Newton} observations \citep{pet02,lon04}
have revealed a puzzling behavior of the iron line in  \object{Mkr 841}. Indeed the
presence of a highly variable, {\it but narrow} iron line feature was
observed in XMM-{\it Newton} 2001, completely at odds with any currently-accepted
interpretation of the line origin \citep{pet02} and required further investigation. A
re-analysis of the XMM-{\it Newton} data by \cite{lon04} proposed that the
line may vary in width rather than in flux. Their interpretation then
invokes local illumination by a flare inducing an hotspot in the inner
disc, which then becomes progressively broadened as the disc rotates. On
the other hand the reflection component, although poorly constrained due
to the lack of high signal to noise in the simultaneous \BS data above 10 keV,
was relatively large (R$>$1) confirming a previous \BS observation
\citep{bia01} . Astonishingly, the continuum shape and flux kept roughly constant
between the two pointings as well as during the total (100 ks) \BS
observation. For a better understanding of the puzzling spectral and
temporal behavior of this source, it has been observed again in 2005 for
a total duration of $\sim$ 75 ks.

We report here on the detailed spectral analysis of the whole XMM-{\it Newton} data
set including one archive observation and the 4 open time pointings done in 2001 and 2005.

\section{The XMM-{\it Newton} data}

\begin{figure*}[t]
\includegraphics[width=\textwidth,angle=0]{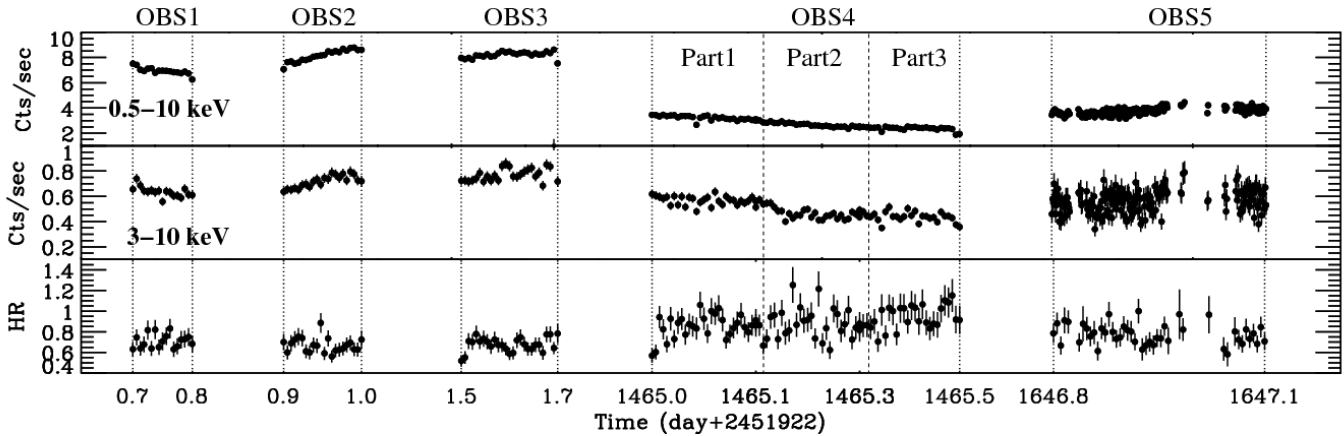}\\
\caption{{\bf Top:} the 0.5-10 keV  X-ray light curve of the different EPIC-pn observations of Mkn
  841. It varies by a factor $\sim$ 5 in 4 years. {\bf Middle:} the 3-10 keV  X-ray light curve of the different EPIC-pn observations of Mkn
  841. It varies by a factor two times lower than the 0.5-10 keV light curve meaning that the broad band variability is dominated by the soft ($<$ 3 keV)
  band one. On the other hand, flux
  variability up to $\sim$50\% is also observed on tens of ks in the  soft and hard X-rays. {\bf Bottom:} Hardness ratio light curve (5-10 keV)/(3-5 keV).}
\label{lcfig}
\end{figure*}

The first \xmm (\citealt{jan01} and references therein) pointing of
 \object{Mkr 841} (denoted OBS 1) was done on the 13th of January 2001 for $\sim$8 ks
 as part of the guaranteed time program. It was immediately
followed by the first open time observation.  Due to
operational contingency, the requested 30 ks were split into two parts,
on the 13th (OBS 2) and the 14th (OBS 3) of January 2001
with $\sim$11 and 13 ks duration time respectively. The two observations
were separated by about 15 hours. The source was re-observed 4 years
later in January 2005. Here again the requested 75 ks
were split into two parts due to strong proton flares
during the observation. About 46 ks of good quality data were extracted
from this pointing (OBS 4) while the 30 ks left were performed in July
2005 (OBS 5). Table \ref{log} gives a summary of the \xmm
pointings with the corresponding dates, observation duration and count rates. \\

In this paper  we
(usually) restrict the analysis to the EPIC-pn data. The EPIC-pn camera 
was always operated in Small
Window mode, with thin aluminum filters to block visible
light. 
The event files were reprocessed from the ODF data files using the
{\it epchain} 
pipeline tasks of the XMM-{\it Newton} Science
Analysis System (SAS version 6.5) and using the most
updated version of the public calibration files. These event files were
then filtered for good time intervals following the ``recipe'' given in
the XMM-{\it Newton} SAS handbook (V2.01 23 July 2004, Sect. 5.2.4). Taking into account the dead time, the filtered event files then contained 5.9, 7.6,  9.4, 30.0 and 12.9 ks of good exposure in the pn detector for OBS 1, 2, 3, 4 and 5 respectively. The final
EPIC-pn 
count rates for each observation are reported in
Tab. \ref{log}. They are always well below the 1\%  pile-up threshold for
both instruments.  Due to the proximity of  the ``Small Window'' edges, the
source spectra and light curves were built from photons detected within a
40 arcsec extraction window centered on the source. X-ray events
corresponding to pattern $\le$ 4 
were selected. The background was estimated within a window
of the same size as the source from an offset position.


In the following, all errors refer to 90\% confidence level for 1
interesting parameter ($\Delta\chi^2$=2.7).

\section{Model-independent analysis}
\subsection{Light curves and hardness ratios}
\label{sectlc}

We have plotted in the upper and middle panel of Fig. \ref{lcfig} the 0.5-10 keV and 3-10 keV EPIC-pn count
rate light curves of the different \xmm observations of Mkn
841 as well as the hardness ratio (5-10 keV)/(3-5 keV) in the lower panel. The time
binning is 500 sec. The total 0.5-10 keV count rate decreases by a factor
$\sim$5 in 4 years while the hardness ratio increases, reaching maximum
values during OBS 4. Between 2001 and 2005, the 3-10 keV count rate shows variations a factor two lower than the total count rate meaning that the broad band count rate variability is dominated by the soft ($<$ 3 keV)
X-ray variability, at least on long time scale.  Smooth soft {\it and}
hard flux variabilities up to $\sim$50\% are also visible on timescale of tens of ks. 


\subsection{RMS}
\label{rmssect}

\begin{figure}[!t]
\includegraphics[width=\columnwidth]{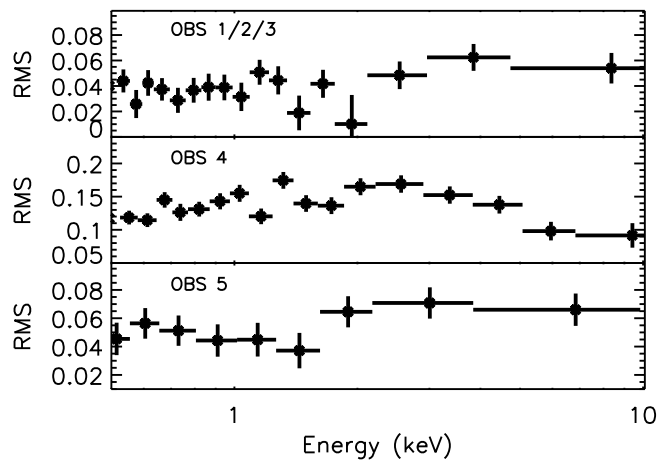}
\caption{RMS spectra of OBS 1/2/3, OBS 4 and OBS 5. }
\label{rms}
\end{figure}

Figure \ref{rms} shows the RMS spectra of the different observations.
The RMS function has been calculated following the procedure detailed in
 \citet{pon04}. We use the 0.5-10 keV energy band. In order to increase the statistics we
have grouped OBS 1, 2 and 3 to produce a single ``2001'' RMS spectrum. The time
binning is of about 1 ks for each variability spectrum while the energy
binning has been chosen in order to have negligible Poisson noise.

 Due to the low degree of variability and the statistics we were not able to produce a spectrum with a large number of energy bins especially at high energy. Nevertheless the different RMS spectra are relatively flat: OBS 1/2/3 and OBS 5 are consistent with a constant  value of $\sim$4 and 5\% respectively. On the other hand the RMS spectrum of OBS 4 is inconsistent with a constant (at more than 99.99\%) with a mean value of 12\%. In this case we note a rough increase of the RMS from 0.5 to about 2-3  keV and then a clear decreases at higher energies down to a few percent. 

From now on, due to the flux and spectral variability observed in OBS 4,  we divide
this observation in 3 parts (noted part1, part2 and part3) of about 15 ks duration each as indicated on Fig. \ref{lcfig}. Each part is analyzed separately.






\subsection{Spectral ratios}
\begin{figure}[!t]
\includegraphics[width=\columnwidth]{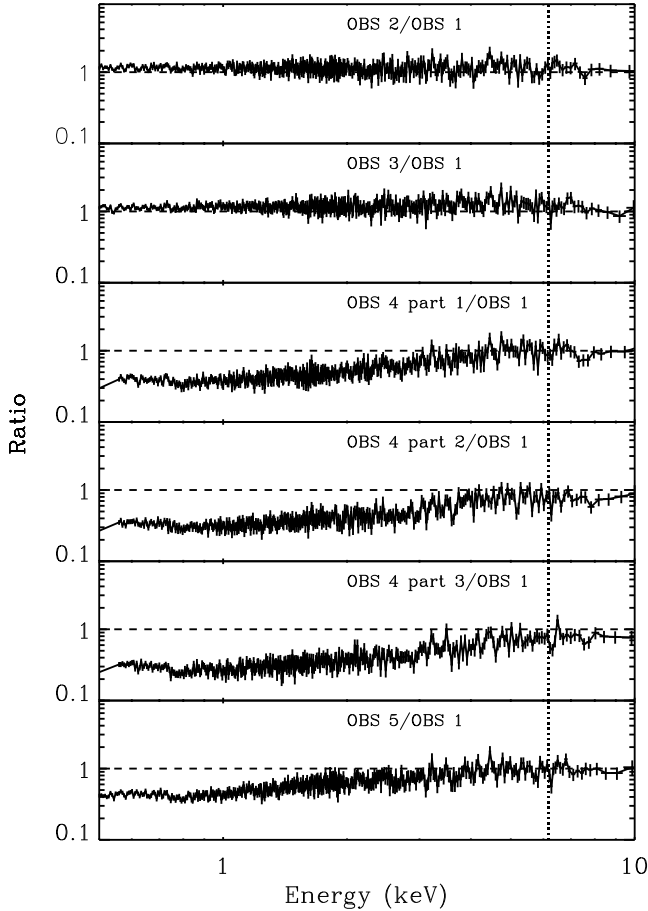}
\caption{Ratios between the different EPIC-pn spectra and the EPIC-pn
  spectrum of OBS1. The dashed line correspond to the position of the 6.4
  keV line in the source frame.}
\label{ratio}
\end{figure}
\begin{figure}[!t]
\includegraphics[width=\columnwidth]{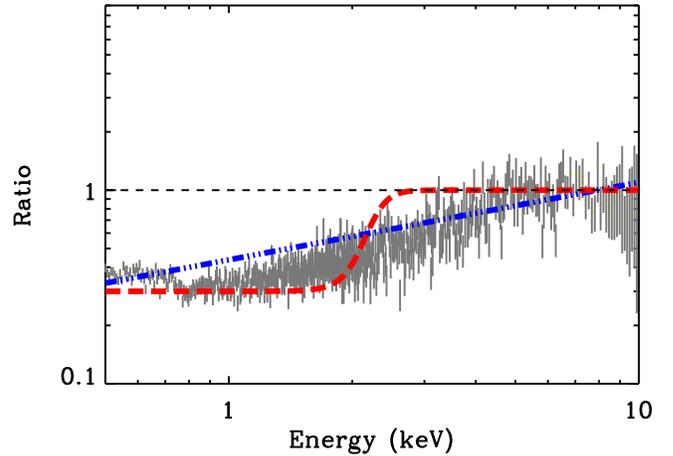}
\caption{Ratio between the EPIC-pn spectrum of the complete OBS 4 and the
  EPIC-pn spectrum of OBS1. The data (white grey lines) have been rebinned in order to have a 5 $\sigma$ confidence level or at least 25 data points per bin. We have over-plotted in this figure, the
  expected variability in two different cases: a pivoting power law
  around 8 keV whose photon index decreases by 0.4 (blue, dot-dashed line) and
  a variable black body peaking at 0.14 keV and varying by a factor 3 in
  flux (red, dashed line).}
\label{ratioobs1-jan05}
\end{figure}


We have plotted in Fig. \ref{ratio} the ratios of the different EPIC-pn
spectra with the EPIC-pn spectrum of OBS1. The binning is such that each
bin has a 5 $\sigma$ confidence level. No strong spectral variations are observed on short ($\sim$ hours) time scale between OBS 1, OBS 2 and OBS 3 or the three parts of OBS 4. However, the 3-8 keV spectral shape clearly hardens between 2001 and
2005, in agreement with the increase of the hardness ratio shown in Fig. \ref{lcfig}.  Above 8 keV the spectra keep roughly  constant
at least in shape.  This spectral variability is more apparent in
Fig. \ref{ratioobs1-jan05} where we have plotted the ratio between OBS 1
and the complete data set of OBS 4. For comparison, we have over-plotted in this figure
the expected variability assuming two different cases: a power
law pivoting around 8 keV whose photon index decreases by 0.4 (dot-dashed line)
and a variable black body peaking at 0.14 keV and varying by a factor 3
in flux (dashed line). These values have been chosen just to illustrate what kind of variability we can expect in such cases. Clearly neither of these simulations is able to reproduce the
complete spectral variability suggesting a more complex behavior.
We can also notice in Fig. \ref{ratio}  some variations of the iron line complex on long (i.e. year like between OBS 1 and part 3 of OBS 4) and short (i.e. hours like between the different parts of OBS 4) time scale.

\section{Spectral analysis}
\subsection{Phenomenological analysis}
\begin{figure}[t!]
\includegraphics[height=\columnwidth,angle=-90]{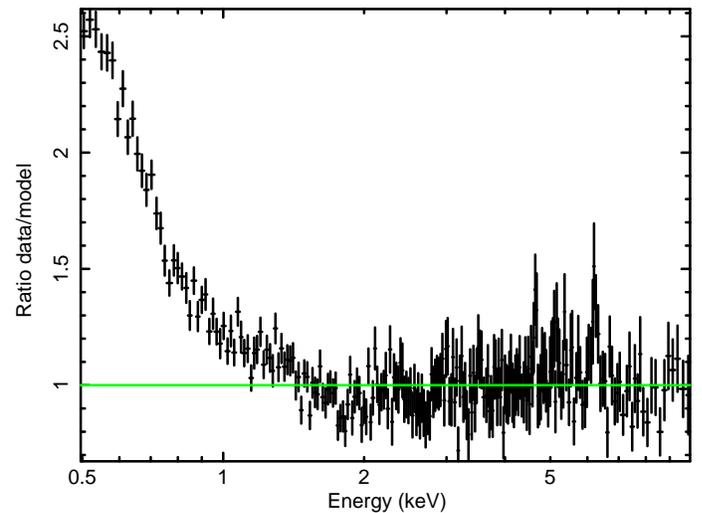}
\caption{Ratio data/model for part 1 of OBS 4. The model is a simple power law
fitted between 3 and 10 keV and extrapolated down to low energies. A strong soft excess and a line near 6 keV are clearly apparent. Note also the presence of a narrow feature near 4.8 keV. It will be discuss in Sect. \ref{redline}. \label{ratioobs4po}}
\end{figure}

Our first step in the spectral analysis was to fit the data with a simple
power law, excluding the data below 3 keV. Figure \ref{ratioobs4po} shows
the ratio between the XMM-{\it Newton}/EPIC-pn data of part 1 of OBS 4 with the corresponding
best fit power law. We clearly observed a strong soft excess below 2
keV and a fluorescent iron line complex near 6.4 keV. 

\begin{table*}
\begin{tabular}{lccccccccccc}
\hline
Obs 
& $\Gamma$ & $E_{FeK_{\alpha}}$ & $\sigma_{FeK_{\alpha}}$ & $F_{FeK_{\alpha}}$ & EW & $kT_{bb}$ & $N_{bb}$ & F$_{0.5-3}$ & F$_{3-10}$& $\Delta\chi^2$& $\chi^2/dof$ \\
&     &  keV        & eV &$10^{-5}$ &eV        &  keV   &           & $10^{-11}$ & $10^{-11}$& \\ 
\hline

1
& 1.81$_{-0.09}^{+0.07}$& 6.25$_{-0.14}^{+0.15}$ & $<$550 & 1.4$_{-0.6}^{+3.2}$ & 90$_{-35}^{+210}$ & 0.20$_{-0.01}^{+0.01}$ & 547$_{-60}^{+70}$  & 1.6 & 1.&17&548/244\\
2
& 1.91$_{-0.08}^{+0.06}$& 6.39$_{-0.05}^{+0.06}$ & 135$_{-55}^{+65}$ & 2.7$_{-0.8}^{+1.0}$ & 170$_{-50}^{+60}$ & 0.18$_{-0.01}^{+0.01}$ & 961$_{-100}^{+110}$ & 1.9 & 1.1&37&510/263\\
3
& 1.95$_{-0.12}^{+0.81}$& 6.54$_{-0.41}^{+0.40}$ &950$_{-570}^{+1020}$ & 5.3$_{-2.0}^{+137.0}$ & 300$_{-200}^{+8400}$ & 0.16$_{-0.01}^{+0.01}$ & 1210$_{-60}^{+70}$ & 1.9& 1.3&32&397/263\\
4 part 1
& 1.43$_{-0.07}^{+0.05}$& 6.44$_{-0.04}^{+0.05}$ &$<$130 & 1.3$_{-0.5}^{+2.5}$ & 90$_{-40}^{+170}$ & 0.13$_{-0.01}^{+0.01}$ & 1996$_{-359}^{+376}$   & 0.8 & 1.0&22&334/273\\
4 part 2
& 1.42$_{-0.04}^{+0.09}$& 5.50$_{-0.43}^{+0.52}$ & 1200$_{-600}^{+2300}$&5.9$_{-4.1}^{+25.5}$ & 420$_{-290}^{+1800}$& 0.14$_{-0.01}^{+0.01}$ & 1021$_{-140}^{+200}$   & 0.6 & 0.8&14&342/266\\
4 part 3
& 1.30$_{-0.06}^{+0.07}$& 6.51$_{-0.03}^{+0.03}$ &$<$80 & 1.6$_{-0.4}^{+2.1}$ & 140$_{-40}^{+80}$& 0.14$_{-0.01}^{+0.01}$ & 1008$_{-130}^{+223}$   & 0.6 & 0.8&37&367/266\\
5
& 1.65$_{-0.06}^{+0.05}$& 6.49$_{-0.07}^{+0.09}$ & 130$_{-50}^{+110}$&1.9$_{-0.8}^{+0.5}$ & 140$_{-60}^{+40}$ & 0.12$_{-0.01}^{+0.01}$ & 3348$_{-680}^{+720}$    & 0.9 & 0.9&25&389/281\\

\hline
\end{tabular}
\caption{Best fit values obtained with a simple model including a power
  law for the continuum, a gaussian for the iron line and a multicolor disc  for the soft excess. The gaussian energy is given in the source frame. The hydrogen column density, not shown in this table, is always consistent with the galactic one, i.e. 2.34$\times$ 10$^{20}$ cm$^{-2}$. The $\Delta\chi^2$ values correspond to the change
  in $\chi^2$ when adding the gaussian component.}
\label{tabfitpobbgau}
\end{table*}

The second step was to include very simple spectral components to
reproduce the observed features. We use a power law for the continuum and
a gaussian for the iron line. 
A more precise analysis of the line complex is done in the next
section. We fit the data above 3 keV first. Then we fix the different
parameters and include the data below 3 keV down to 0.5 keV and we add a
simple multicolor accretion disc component ({\sc discbb} in {\sc xspec}) to model the soft excess. The best fit
parameter values obtained with this method for the continuum, the  multicolor disc and the line are reported in Table \ref{tabfitpobbgau}.  

\begin{figure*}[t!]
\begin{tabular}{cc}
\includegraphics[width=\columnwidth]{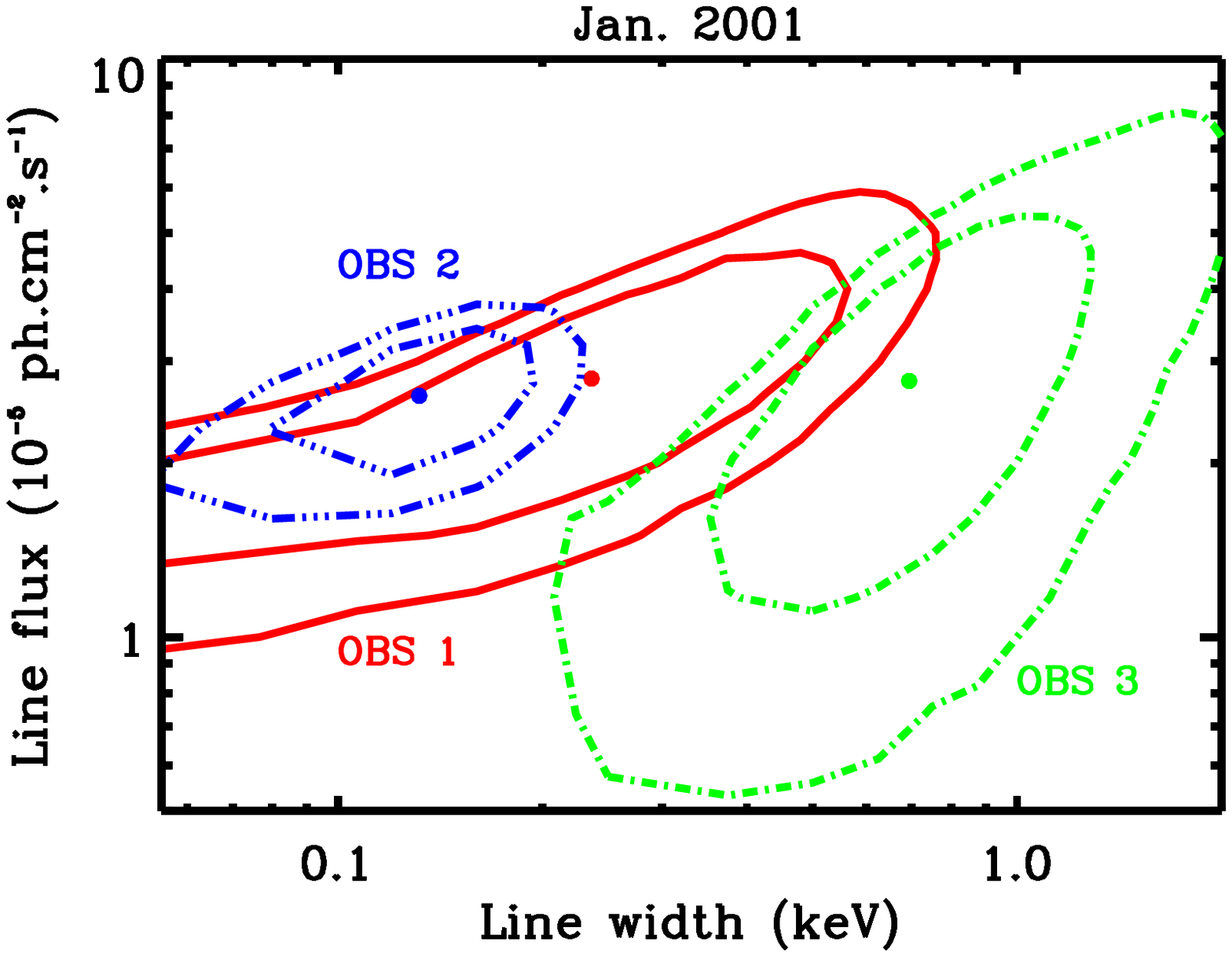}&
\includegraphics[width=\columnwidth]{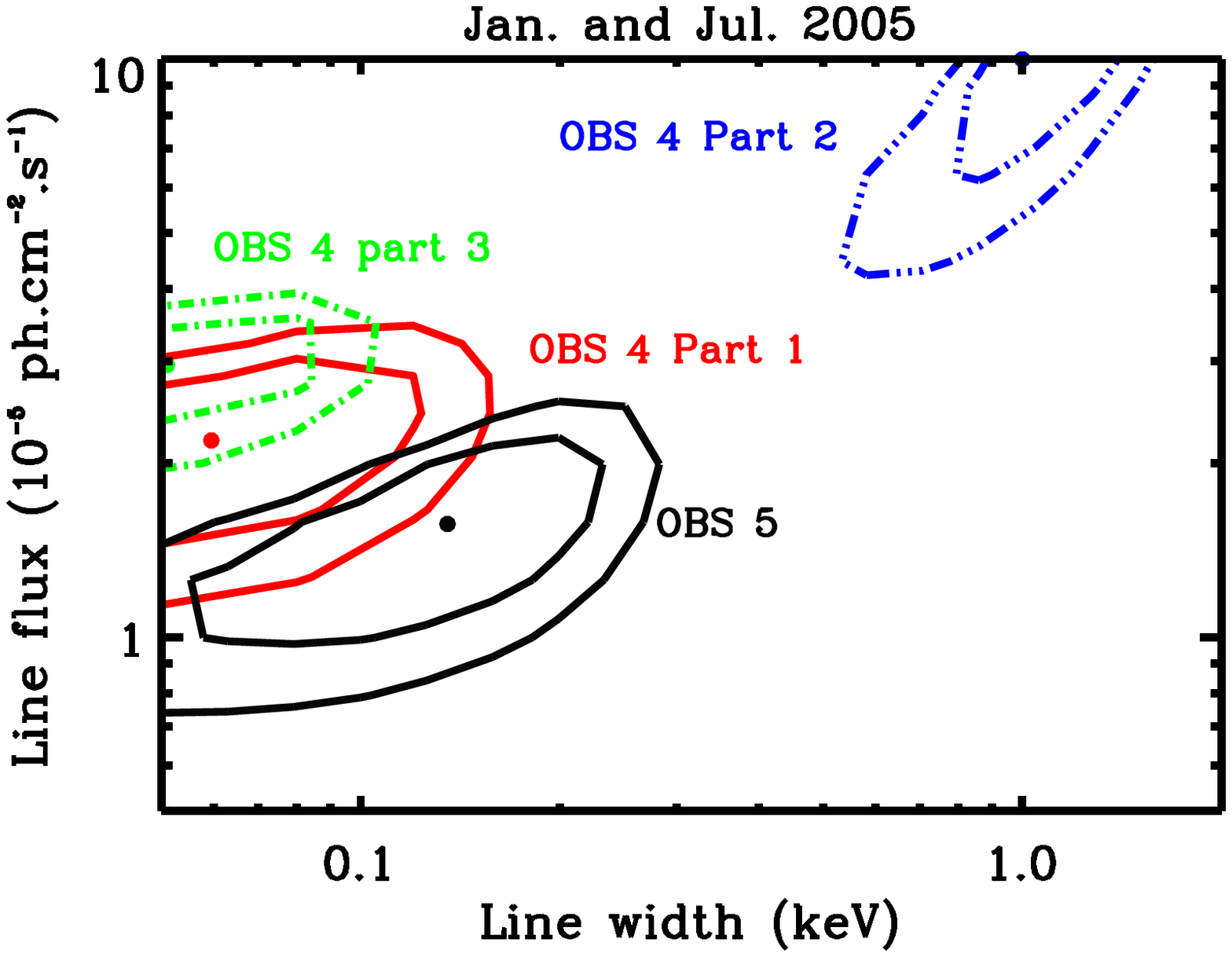}\\
\end{tabular}
\caption{Contour plots (68 and 90 \%) of the line width vs. line flux
  obtained for the 3 observations of Jan 2001
  (left) and the 3 parts of Jan. 2005 as well as Jul. 2005 (right). The model
  includes a simple power law + gaussian line. 
  \label{contplot}}
\end{figure*}

The best fits are clearly never satisfactory and
large discrepancies are present especially below 3 keV due to the bad black body approximation for the soft excess. Nevertheless several
remarks, weakly affected by the goodness of the fit, can still been made:

%
%
%
%
%
%
%
%
%


\begin{itemize}
\item The photon index reaches values as small as 1.3 during OBS 4 where the flux is the lowest.
\item The flux variation in the soft band ($<$ 3 keV) between 2001 and Jan 2005 is more than a factor 2 larger than the flux variation above 3 keV . 
\item The iron line width appears variable on relatively short time scale ($<$ 15 ks) e.g. between part 1 and part 2 of OBS 4. 
\end{itemize}
The spectral variability agrees with the results shown previously with model-independent methods and suggest a spectral pivot at high energy. However the presence of a simple power law continuum is unlikely given the unusually (for a Seyfert galaxy) hard spectral index of OBS 4. More physical models are discussed in Sect. \ref{physmodel}. Concerning the iron line, the 2005 observations confirm the apparent line variability already observed in 2001 \citep{pet02,lon04}. 
The best fits reported in Table \ref{tabfitpobbgau} indicate the presence of a narrow line (compared to the EPIC-pn resolution)  except OBS
  3 and part 2 of OBS 4 where a broad component is preferred. The contour
plots (at 68 and 90 \% confidence level) of the line flux vs. line width
for the different observations are plotted in Fig. \ref{contplot}  for 2001 (left plot) and 2005 (right plot). 
The line variability is discussed in more detail in the next section. 

\begin{figure*}[!t]
\begin{center}
\includegraphics[width=0.7\textwidth]{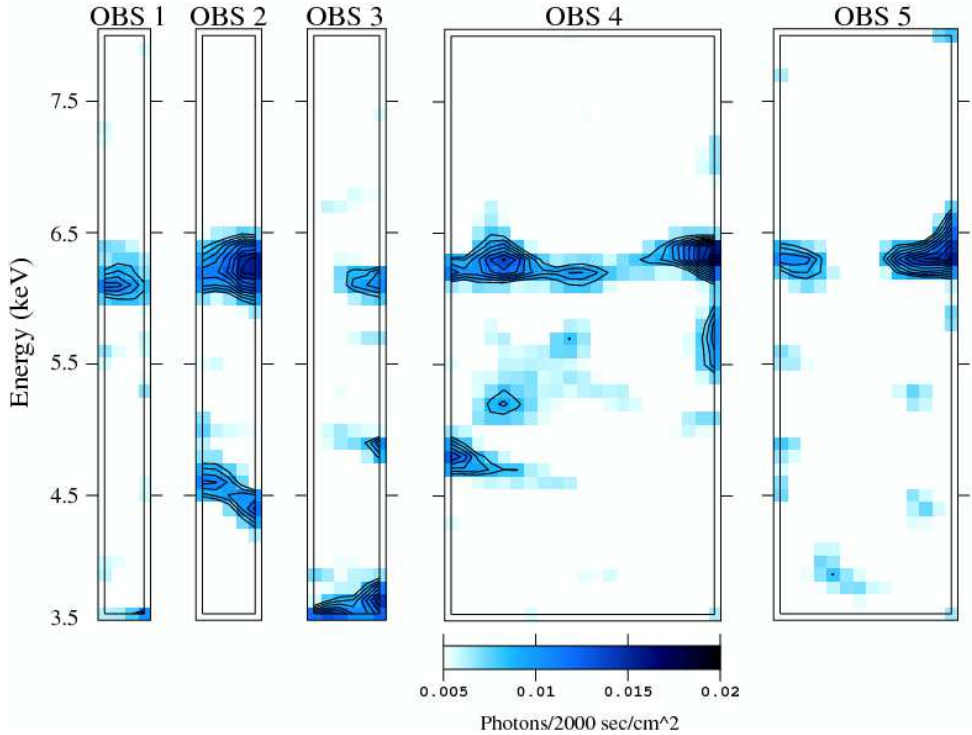}
\caption{Excess map on time--energy plane of the different XMM observations. The energy scale is in the lab frame. \label{mapexcess}}
\end{center}
\end{figure*}
To decrease the number of degrees of freedom, we choose to  follow the
fitting procedure of \citet{lon04}. We fit  together observations with roughly the same underlying continuum shape (cf. Tab. \ref{tabfitpobbgau}) , e.g. the three pointings of 2001, on the one hand, and the different part OBS 4, on the other, keeping  the power law continuum  constant in shape but not in flux between the different spectra. OBS 5 is analyzed as a single observation.

\subsection{The iron line complex}
\label{ironline}

\subsubsection{Rapid variability}

The rapid ($\sim$hours time scale) variability of the line complex seems to be a common characteristic of this source. For a better
visualization of this variability we have plotted in Fig. \ref{mapexcess}
the excess map of the complete set of the XMM-{\it Newton}/EPIC-pn observations.
This map has been obtained following the method of \cite{iwa04} and \cite{tomb07} applied for NGC 3516 and NGC 3783 respectively. 
We use resolutions of 2 ks in time and  100 eV in energy. Each temporal slice represents the residuals obtained when fitting each 2 ks spectrum with a power law between 3.5 and 8 keV but ignoring data between 4 and 7 keV. This map is useful to reveal narrow features but signatures of broad ones can be mixed up with the underlying continuum.  A narrow feature close to 6.4 keV is
clearly present during most of the different pointings, but it seems to
disappear from time to time on a very short time scale ($\sim$ ks) like in OBS 3 or in the middle of OBS 4 and
OBS 5 .
\begin{figure}[b]
\includegraphics[width=\columnwidth]{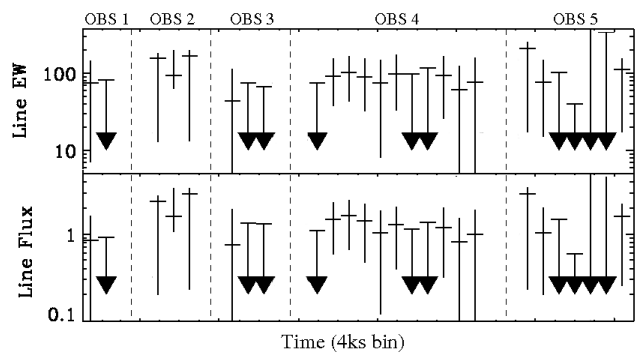}
\caption{Equivalent width (top, in eV) and flux (bottom, in 10$^{-5}$ ph cm$^{-2}$ s$^{-1}$) light curves of a narrow ($\sigma$=0 eV) gaussian line whose energy is fixed to 6.4 keV. We use a 4 ks time binning \label{lcew}}
\end{figure}

To check if the narrow line component variability is real or not  we produce the light curves of its flux and equivalent width. They are plotted in Fig. \ref{lcew} with a 4 ks binning time scale. The model includes a power law and a narrow ($\sigma=$ 0 eV) gaussian line with energy fixed at 6.4 keV. We fit the data between 3 and 10 keV. From time to time the narrow line is poorly detected with only upper limit on its flux and EW. However both are  consistent with a constant (at more than 98\%) from 2001 to 2005. The fact that the EW is also constant  is not surprising given the relative constancy of the underlying continuum near 6.4 keV (see Fig. \ref{ratio}). A constant narrow  line flux suggests the presence of remote reflection and it is discussed in the next section. If this interpretation is correct, the apparent variability shown by the excess map may reveal changes in either the continuum or any broad line emission underlying rather than changes of the narrow line itself. This is discussed in Sect. \ref{releff}.

\subsubsection{Remote reflection?}
\label{remoteref}
\begin{figure}[t!]
\includegraphics[width=\columnwidth]{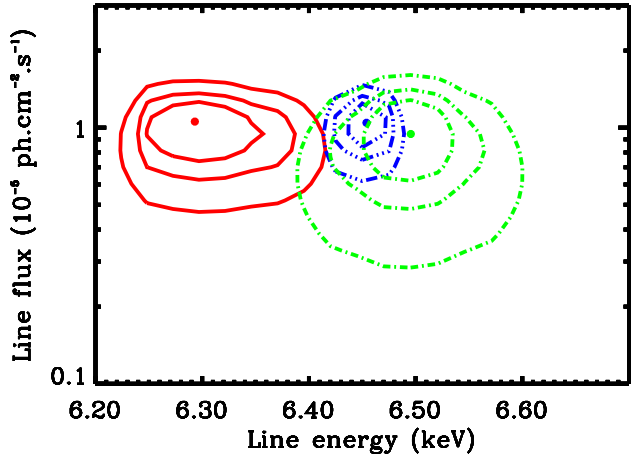}
\caption{Contour plots (68, 90 and 99\%) line flux vs. line energy (source frame) of the narrow ($\sigma=$0 eV) line component added to mimic emission from remote material. OBS 1/2/3: red/solid contour, OBS 4 part 1/2/3: blue/dot-dot-dot-dashed contours, OBS 5: green/dot-dashed contours. \label{contlinetorus}}
\end{figure}
The presence of a roughly constant narrow line component, as shown in Fig. \ref{lcew},  suggests the presence of neutral reflection from remote material. Such reflection may come from the outer part of the accretion disc or even farther away from e.g.  the dusty torus  surrounding the nucleus as expected in the unification framework of AGNs. 

In Tab. \ref{linetorus} we report the best fit parameters of the narrow ($\sigma$=0 eV) gaussian line added to the 3-10 keV power law best fit of the three observations of 2001 keeping the power law flux free to vary between OBS 1, 2 and 3. We apply the same method for the three parts of OBS 4. OBS 5 is analyzed normally as a single observation. The contour plots of the narrow line flux vs.  line energy are plotted in Fig. \ref{contlinetorus}. The line fluxes obtained in the different observation periods are consistent with each other, in agreement with a roughly constant remote reflection. Surprisingly enough, while the best fit energy of the line  is consistent with a slightly ionized iron line in 2005 (the 90\% confidence level energy range being in between 6.43 and 6.47 keV  for OBS 4 and in between 6.43 and 6.54 keV for OBS 5) it is smaller than and marginally consistent with 6.4 keV (at less than 10\% confidence level) in 2001. 
\begin{table}
\begin{center}
\begin{tabular}{lcc}
\hline
Obs & $E_{FeK_{\alpha}}$ & $F_{FeK_{\alpha}}$ \\ 
\hline
1/2/3 & 6.31$_{-0.10}^{+0.05}$ & 10.2$_{-3.1}^{+2.7}$ \\ 
4 part 1/2/3 & 6.45$_{-0.02}^{+0.02}$ & 10.4$_{-2.3}^{+2.3}$ \\ 
5 & 6.49$_{-0.06}^{+0.04}$ & 10.2$_{-4.5}^{+3.1}$ \\ 
\hline
\end{tabular}
\caption{Best fit parameters of the narrow ($\sigma=$0 eV) gaussian line added to mimic emission from remote material. The line energy is given in the source frame and the line flux in $10^{-6}$ ph cm$^{-2}$ s$^{-1}$. 
}
\label{linetorus}
\end{center}
\end{table}

In this last case, the detected line may be the signature of a slightly redshifted narrow iron line. This interpretation will be discussed in more detail in Sect. \ref{redline}. If it is correct, the presence of remote reflection should still be tested for.  We check the presence of a second narrow line component
fixing the parameters of the first line to their best fit values. The best fit parameters for this second narrow line  are $E_{line}$=6.48$_{-0.06}^{+0.12}$ keV and $F_{line}$=5.7$_{-3.4}^{+2.2} \times 10^{-6}$ ph cm$^{-2}$ s$^{-1}$. These parameters are now  in better agreement with those obtained for OBS 4 and OBS 5. 

\subsubsection{Relativistic effects?}
\label{releff}
\begin{table*}
\begin{center}
\begin{tabular}{lcccccccccc}
\hline
Obs &  $\Gamma$ & $q$ & $r_{in}$ &  & & $E_{FeK_{\alpha}}$ & EW &  &  & $\chi^2/dof$ \\
\hline
& & & &\multicolumn{2}{c}{OBS 1}&\multicolumn{2}{c}{OBS 2}&\multicolumn{2}{c}{OBS 3}& \\
1/2/3 &1.85$_{-0.04}^{+0.04}$& $>$2.5 & 270$_{-70}^{+160}$ & 6.41$_{-0.10}^{+0.11}$& 100$_{-50}^{+60}$&  6.50$_{-0.06}^{+0.06}$ & 130$_{-40}^{+60}$& 7.06$_{-0.10}^{+0.11}$  & 70$_{-40}^{+50}$& 392/424\\
\hline
& & & &\multicolumn{2}{c}{part 1}&\multicolumn{2}{c}{part 2}&\multicolumn{2}{c}{part 3}& \\
4 part 1/2/3&1.40$_{-0.04}^{+0.04}$& $>$3.9 & 11$_{-2}^{+2}$ & 6.35$_{-0.20}^{+0.16}$  & 180$_{-110}^{+90}$& 6.21$_{-0.16}^{+0.10}$  & 220$_{-90}^{+90}$& 6.44$_{-0.11}^{+0.08}$& 310$_{-110}^{+110}$&424/442 \\
\hline
5 &1.65$_{-0.06}^{+0.06}$& $>$5.0 & $<$20 & & &6.92$_{-0.51}^{+0.37}$  & $230_{-140}^{+140}$&   & & 167/158\\
\hline
\end{tabular}
\end{center}
\caption{Best fits parameter values obtained with a power law + relativistic line ({\sc diskline}) + narrow ($\sigma$=0 eV) gaussian line model, fitting the data above 3 keV. OBS 1, 2 and 3 are fitted simultaneously keeping the power law index $\Gamma$, the disc emissivity power law index $q$  and the inner disc radius $r_{in}$ constant between the different observations. We apply the same procedure for part 1, 2 and 3 of OBS 4.  The line energy is  given in the source frame, $r_{in}$ is in unit of $r_g$ and is larger than 6 (Schwarzschild metric). The inclination angle is fixed to 30 deg and the outer radius to 1000 $r_g$. \label{tabfitsimultdiskline}}
\end{table*}

\begin{table*}
\begin{center}
\begin{tabular}{lccccccccccc}
\hline
Obs &  $\Gamma$ &   & & & $N_{h}$ & $\log{\xi}$ & $\sigma$ &  & & &$\chi^2/dof$ \\
\hline
& &  \multicolumn{3}{c}{OBS 1}&\multicolumn{3}{c}{OBS 2}&\multicolumn{3}{c}{OBS 3}& \\
1/2/3 & 2.14$_{-0.04}^{+0.20}$&35.0$_{-20.2}^{+4.9}$&3.4$_{-0.1}^{+0.1}$&$>$0.3&43.4$_{-7.5}^{+4.8}$&3.5$_{-0.0}^{+0.1}$&$>$0.3&$>$43.6&3.4$_{-0.1}^{+0.1}$&$>$0.3&423/423\\\\
\hline
& & \multicolumn{3}{c}{part 1}&\multicolumn{3}{c}{part 2}&\multicolumn{3}{c}{part 3}& \\
4 part 1/2/3&1.75$_{-0.15}^{+0.14}$&25.9$_{-}^{+}$&3.3$_{-1.0}^{+0.3}$&0.2$_{-0.1}^{+0.2}$&8.6$_{-1.7}^{+5.8}$&2.4$_{-0.2}^{+0.3}$&$_{-}^{+}$&$>$29.5&3.4$_{-0.7}^{+0.1}$&0.3$_{-0.1}^{+0.2}$&423/441\\
\hline
5 &1.76$_{-0.06}^{+0.13}$& & $<$8.0& & &$<$2.5 & & &0 & &171/159\\
\hline
\end{tabular}
\end{center}
\caption{Best fit parameter values obtained with a power law + relativistic smeared absorption  ({\sc swind}) + narrow ($\sigma$=0 eV) gaussian line model, fitting the data above 3 keV. OBS 1, 2 and 3 are fitted simultaneously keeping the power law index $\Gamma$  constant between the different observations. We apply the same procedure for part 1, 2 and 3 of OBS 4.   The absorbing column density $N_{h}$ is in unit of $10^{22}$cm$^{-2}$. \label{tabfitsimultswind}}
\end{table*}

\begin{figure}[b!]
\includegraphics[width=\columnwidth,angle=0]{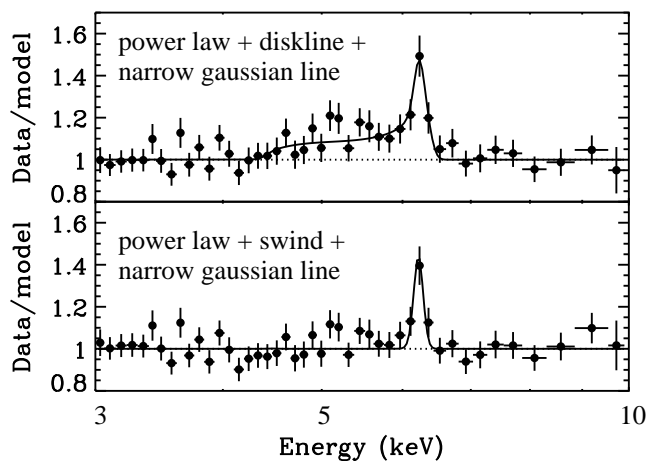}
\caption{Ratio data/model for OBS 4. {\bf Top:} The model is a simple power law + {\sc diskline} + narrow ($\sigma$=0 eV) gaussian line
fitted between 3 and 10 keV. We ignore the best fit line components (over-plotted in solid line) to produce this ratio.  {\bf Bottom:} The model is a simple power law + a partially ionized absorbing material with large velocity shear + narrow ($\sigma$=0 eV) gaussian line. We also ignore the best fit line component (over-plotted in solid line) to produce this ratio. The presence of a broad line component is strongly reduced and only the narrow component is present (see Sect. \ref{ironline}).\label{broadline}}
\end{figure}
If we believe in the constancy of the narrow line flux, the variability shown by the map excess (Fig. \ref{mapexcess}) may result from variations in some other underlying component. The most common explanation of such variability is  the presence of a broad and rapidly variable line component. This is also illustrated in the upper panel of Fig. \ref{broadline} where
we have plotted the ratio of the complete OBS 4 data spectrum and the
best fit power law + {\sc diskline} + narrow ($\sigma$=0 eV) gaussian line model. { The {\sc diskline}  model \citep{fab89} is a relativistic accretion disc line model around a static (Schwarzschild) black hole whose parameters are the line energy, the inner and outer disc radii, the power law index of the disc emissivity law and the disc inclination}.

This ratio has been  obtained by fitting the data above 3 keV  and fixing the inclination to 30 degrees.  This figure is then obtained after setting the line normalizations to zero. The presence of a complex line profile is clearly visible with a broad component down to $\sim$ 4.5 keV. The solid line over-plotted in this figure corresponds to the {\sc{diskline}} + narrow line shape.

The corresponding best fit values are a power law photon index $\Gamma=1.52_{-0.05}^{+0.06}$, a {\sc diskline}  energy $E_{FeK_{\alpha}}=6.12_{-0.57}^{+0.46}$ keV, a line EW of $280_{-110}^{+130}$ eV, an inner disc radius $r_{in}=10.2_{-4.1}^{+13.9}$ $r_g$ and a disc emissivity power law index $q>2.4$. The accretion disc outer radius is fixed to 1000 $r_g$. The fit is very good with a $\chi^2$/dof=147/165, suggesting that relativistic effects could indeed be a good explanation of the line profile. We apply this model to the different observations  of  \object{Mkr 841}. During the simultaneous fit of  OBS 1, OBS 2 and OBS 3 not only  $\Gamma$ but also $q$ ad $r_{in}$ are kept constant between the different observations. We let the power law normalization as well as the {\sc diskline} line flux and energy free to vary. We apply the same procedure for the three parts of OBS 4. Concerning OBS 5, it is analyzed normally as a single observation. We fix the inclination angle to 30 deg and the outer radius to 1000 $r_g$.  We obtain good fits in all cases. The corresponding best fit parameter values are reported in Tab. \ref{tabfitsimultdiskline}.

An alternative to the broad line component could be the presence of ionized warm absorption, smeared by relativistic effects, whose imprints on the underlying continuum could mimic a broad line component (e.g. \citealt{ree04}). Such a model has also been suggested recently in the literature to produce the soft excess in AGNs and will be discussed more precisely in the next section, but we test this model here to see its impact on the iron line profile. We fit the 3-10 keV data with a simple power law and a partially ionized absorbing material with large velocity shear ({\sc swind} model originally proposed by \citet{gie04} and updated by \citet{gie06}). We also add a narrow gaussian line to mimic the remote reflection.  For comparison with the {\sc diskline} model, we have plotted in the lower panel of Fig. \ref{broadline} the ratio data/model for the complete OBS 4 data spectrum. The presence of a broad line component is strongly reduced and only a narrow component is observed.  The corresponding best fit values are a power law photon index $\Gamma=1.9_{-0.3}^{+0.5}$,  a column density and ionization parameter of the absorbing material $N_h> 36 \times 10^{22}$cm$^{-2}$ and $\xi=3080_{-2100}^{+1220}$ and a gaussian velocity smearing $\sigma=0.19_{-0.05}^{+0.08}$. The fit is statistically acceptable with a  $\chi^2$/dof=176/168. While it is larger than the {\sc diskline} fit, both are statistically equivalent following the F-test . We apply this model to the different XMM-{\it Newton} observations. We obtain good fits in all cases. The corresponding best fit parameter values are reported in Tab. \ref{tabfitsimultswind}.

In conclusion, the data are consistent with the presence of a constant narrow line, potential signature of a remote reflection. Then the apparent line variability  shown by the excess map (Fig. \ref{mapexcess}) may result more likely from slight changes of the underlying continuum. We have tested two possible origins of such variability. It could be due to the presence of a rapidly variable broad iron line component. It could also result from variable relativistically smeared absorption features. Both cases require strong relativistic effects to agree with the data.

\label{line}



\subsection{A broad band physical analysis}
\label{physmodel}
Up to now, our spectral analysis was relatively phenomenological, using simple components to fit the different spectral features present in the data (soft excess, broad and narrow line) and focusing on the high ($>$ 3 keV) energy range. The next step is the use of more physical and consistent models on the total energy range of the EPIC-pn instrument.


Recent studies suggest that relativistically blurred (photoionized) reflection
from the accretion disc could be an appealing explanation for the presence of
strong soft excesses in AGNs \citep[e.g.][]{cru06}. Moreover, in
the cases in which a broad Fe line is clearly detected (such as in
MCG--6-30-15) the model is very robust because the soft excess and broad
Fe line are fitted self--consistently with the same relativistically
blurred reflection model. \cite{cru06} already
applied this model to OBS 3  with success. 
As said in the previous section, an alternative explanation for the origin of the soft excess assumes the presence of absorption features. Since the soft excess  does not show strong edges nor absorption lines,  strong velocity gradients are needed in the absorbing medium  to  smear out these features \citep{gie04,sob07,sch06,gie06}. A disc wind could produce such spectral signatures. However, this model does not directly explain the presence of a broad line feature and, if present, it has to be produced by another component. On the other hand, as shown in the previous section,  ionized absorption  may have some impacts on the continuum spectral shape near 6 keV and consequently on the observed line  broadness (cf. Fig. \ref{broadline}).  

Our XMM-{\it Newton} data of  \object{Mkr 841} appear to be well adapted to test both interpretations. 


\subsubsection{The fitting procedure}

\begin{table*}
\begin{center}
\begin{tabular}{lccccccc}
\hline
Obs. 
 & N$_h$ & $\Gamma$ &   $N_{neutral}$ &$N_h^{WA}$ & $\xi^{WA}$ & $\Delta\chi^2$ & $\chi^2$/dof\\
 & 10$^{20}$ cm$^{-2}$& &  $\times$10$^5$& 10$^{22}$ cm$^{-2}$ & & & \\
 \hline
 1/2/3 
 &$<$2.8 & 2.45$_{-0.04}^{+0.05}$  &7.9$_{-1.5}^{+2.0}$ &1.1$_{-0.4}^{+0.1}$& $>$900 &30 &783/788\\
 4 - 1/2/3 
 & 7.4$_{-0.9}^{+1.1}$& 1.59$_{-0.02}^{+0.01}$ & 6.4$_{-1.5}^{+1.4}$ & 2.7$^{-0.5}_{+0.8}$ & 1500$^{-350}_{+470}$ & 72 &865/809\\
 5  & 6.5$_{-1.6}^{+4.0}$& 2.01$_{-0.09}^{+0.05}$ &  5.8$_{-2.3}^{+2.4}$&0.4$_{-0.4}^{+0.3}$& 50$_{-10}^{+10}$ & 75 &303/278\\
 \hline
\end{tabular}
\end{center}
\caption{Best fit parameter values for the continuum, the neutral reflection and the WA obtained with the relativistically blurred ionized reflection \r
  model. $N_h$ is the neutral hydrogen column density,  $\Gamma$ the continuum photon index, $N_{neutral}$ the neutral reflection normalization, $N_h^{WA}$ the ionized  hydrogen column density of the WA  and $\xi^{WA}$ its ionization parameter .$\Delta\chi^2$ is the fit improvement with the addition of the warm absorber. The corresponding best fit parameter values of the reflecting material are reported in Tab. \ref{tabkdbwa}.\label{tabcontinuumkdbwa}}
\end{table*}

\begin{table*}
\begin{center}
\begin{tabular}{lccccccc}
\hline
Obs. 
 & N$_h$ & $\Gamma$ &   $N_{neutral}$ &$N_h^{WA}$ & $\xi^{WA}$ & $\Delta\chi^2$ & $\chi^2$/dof\\
 & 10$^{20}$ cm$^{-2}$& &  $\times$10$^5$& 10$^{22}$ cm$^{-2}$ & & & \\
 \hline
 1/2/3  & $<$2.6 & 2.22$_{-0.01}^{+0.03}$ & 10.7$_{-2.0}^{+2.0}$&0.10$_{-0.03}^{+0.01}$ &60$_{-20}^{+30}$ &15 &824/787\\
 4 - 1/2/3 &$<2.8$ &1.90$_{-0.04}^{+0.04}$ & 7.7$_{-1.5}^{+1.6}$& 0.2$_{-0.1}^{+0.1}$& 20$_{-10}^{+15}$&106 &843/808\\
 5 & $<$6.0&1.96$_{-0.02}^{+0.02}$ & 7.0$_{-2.5}^{+2.5}$& 0.4$_{-0.1}^{+0.1}$&40$_{-10}^{+20}$ & 70& 315/279\\
\hline
\end{tabular}
\end{center}
\caption{Best fit parameter values for the continuum, the neutral reflection and the WA obtained with the relativistically smeared ionized absorption \a
  model. The parameters definition is the same as in Tab. \ref{tabcontinuumkdbwa}. The corresponding best fit parameter values of the absorbing material are reported in Tab. \ref{tabswindwa}. \label{tabcontinuumswindwa}}
\end{table*}

\begin{table*}
\begin{center}
\begin{tabular}{lccccccccccc}
\hline
Obs.  &  & $\xi_{REF}$ & & $q$ & $r_{in}$ & Refl. frac. \\
 & & & & & $r_g$ \\
 \hline
 & OBS 1 & OBS 2 & OBS 3 & & \\ 
 1/2/3  & 130$_{-10}^{+10}$& 155$_{-10}^{+20}$& 100$_{-10}^{+10}$ &$>$8.6 & $<$1.4 &30\%\\
 \hline
 & part 1 & part 2 & part 3 & & \\ 
 4 - 1/2/3 
  & 315$_{-10}^{+10}$ & 325$_{-10}^{+10}$ & 200$_{-40}^{+40}$ &
 3.7$_{-0.2}^{+0.2}$ & $<$2.5& 38\%\\
 \hline
 5 
  & & 80$_{-20}^{+40}$& &4.3$_{-0.4}^{+0.4}$ & $<$2.7 & 43\%\\
 \hline
\end{tabular}
\end{center}
\caption{Best fit parameter values characterizing the ionized reflecting material in the \r model. $\xi_{REF}$ is the ionization  parameter  of the ionized reflection component, $q$ the disc emissivity power law index and  $r_{in}$ the inner radius  of
the reflecting accretion disc. The reflection fraction is the ratio of the ionized reflection flux in the 0.1-1000 keV band divided by the total 0.1-1000 keV flux  of the best fit model.\label{tabkdbwa}}
\end{table*}

\begin{table*}
\begin{center}
\begin{tabular}{lccccccccc}
\hline
Obs. 
  &  & $N_h^{ABS}$& & & $\log(\xi_{ABS})$ & & & $\sigma_{ABS}$ & \\
 & & 10$^{22}$ cm$^{-2}$& & & & & & c &   \\
  \hline
 &  OBS 1 & OBS 2 & OBS 3 & OBS 1 & OBS 2 & OBS 3 & OBS 1 & OBS 2 & OBS 3 \\ 
 1/2/3 
 & 23.1$_{-0.7}^{+2.3}$& 24.4$_{-1.5}^{+1.5}$& 14.3$_{-1.1}^{+1.3}$ & 3.22$_{-0.03}^{+0.05}$& 3.27$_{-0.03}^{+0.04}$& 3.01$_{-0.03}^{+0.02}$& $>$0.4 & 0.36$_{-0.04}^{+0.05}$& $>$0.45\\
 \hline
 &  part 1 & part 2 & part 3 & part 1 & part 2 & part 3 & part 1 & part 2 & part 3 \\ 
 4 - 1/2/3 
  & 15.5$_{-1.3}^{+2.0}$& 20.4$_{-1.3}^{+3.3}$& 17.5$_{-2.0}^{+1.5}$& 2.90$_{-0.05}^{+0.04}$& 2.99$_{-0.05}^{+0.06}$& 2.90$_{-0.05}^{+0.05}$& 0.25$_{-0.03}^{+0.03}$ & 0.29$_{-0.03}^{+0.04}$ & 0.26$_{-0.03}^{+0.04}$ \\
 \hline
5 
 & & 16.7$_{-8.6}^{+5.9}$& & & 3.22$_{-0.18}^{+0.16}$&  & & 0.30$_{-0.04}^{+0.10}$& \\
 \hline
\end{tabular}
\end{center}
\caption{Best fit parameter values characterizing the absorbing material in the \a model. $N_h^{ABS}$ and $\xi_{ABS}$ are the column density  and ionization parameter  of the absorbing wind and $\sigma_{ABS}$ the gaussian velocity smearing. \label{tabswindwa}}
\end{table*}
 


Our base line model has the
following components: 1) a neutral absorption free to vary above the
galactic value,  2) a cut-off power law continuum (the high energy cut-off
being fixed to 300 keV) and 3) a
neutral reflection,  if needed, to reproduce the narrow line component.
For the neutral reflection we use the
tables of the Ross \& Fabian code \citep{ros05} and we fix the ionization parameter to 1 to account for a neutral medium and the illuminating power law continuum to 1.9. This value corresponds to the average 2-10 keV X-ray photon index of Seyfert galaxies (e.g. \citealt{mat01}), but the fit results do not depend significantly on this parameter. 
The free parameters of the first three components of our model are the hydrogen column density, the cut--off continuum power law photon index and normalization and the normalization of the neutral reflection component.

Then we add a fourth component to reproduce the soft excess i.e.  either a relativistically blurred ionized reflection  or a relativistically smeared ionized absorption (noted \r and \a respectively in the following). For the ionized reflection, we also use the
tables of the Ross \& Fabian code \citep{ros05}. The blurring is done by convolving the reflection spectrum with a
Laor profile ({\sc kdblur} kernel). { The Laor model is a relativistic accretion disc line model around a maximally rotating (Kerr) black hole \citep{lao91}}. The corresponding free parameters of \r are the inner radius $r_{in}$ of
the reflecting accretion disc, the disc emissivity power law 
index $q$, the  normalization and the ionization  parameter $\xi_{REF}$ of the ionized reflection component. The iron abundance is fixed to the
solar one for the computation of the reflection,  the disc outer radius to 1000 Schwarzschild radii and  the inclination angle to 30 degrees.

Concerning  \a we use again the {\sc swind} model developed by \citet{gie06} for {\sc xspec}. The free parameters are the column density $N_h^{ABS}$ and ionization parameter $\xi_{ABS}$ of the absorbing wind and the gaussian velocity smearing $\sigma_{ABS}$.

We apply these different models to the 0.5-10 keV energy range of the EPIC-pn detectors. Like in the previous section, we analyzed OBS 1, 2, 3 simultaneously. We keep all the parameters constant between the different data sets except the normalization of the cut-off power law continuum and, for {\sc REF} the normalization and ionization parameters of the ionized reflection, and, for {\sc ABS}, the column density and ionization parameter  of the absorbing wind as well as the gaussian velocity smearing. We apply the same procedure for the three parts of OBS 4.  Then we have 14 free parameters for \r and 15 for {\sc ABS}. OBS 5 is analyzed as a single observation with 8 and 7 free parameters for \r and \a respectively.

\begin{figure*}[th!]
\begin{tabular}{cc}
\includegraphics[width=\columnwidth]{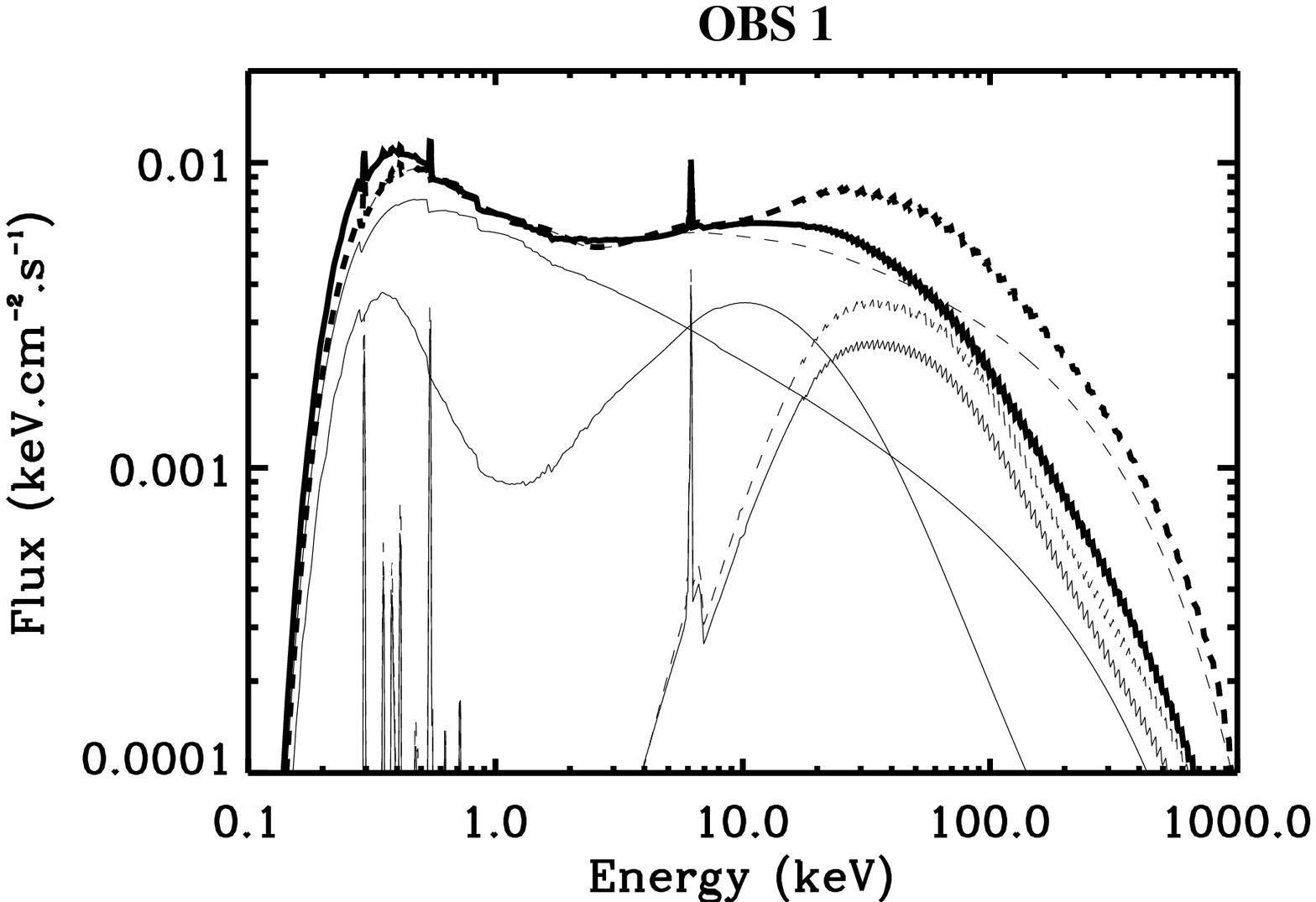}&
\includegraphics[width=\columnwidth]{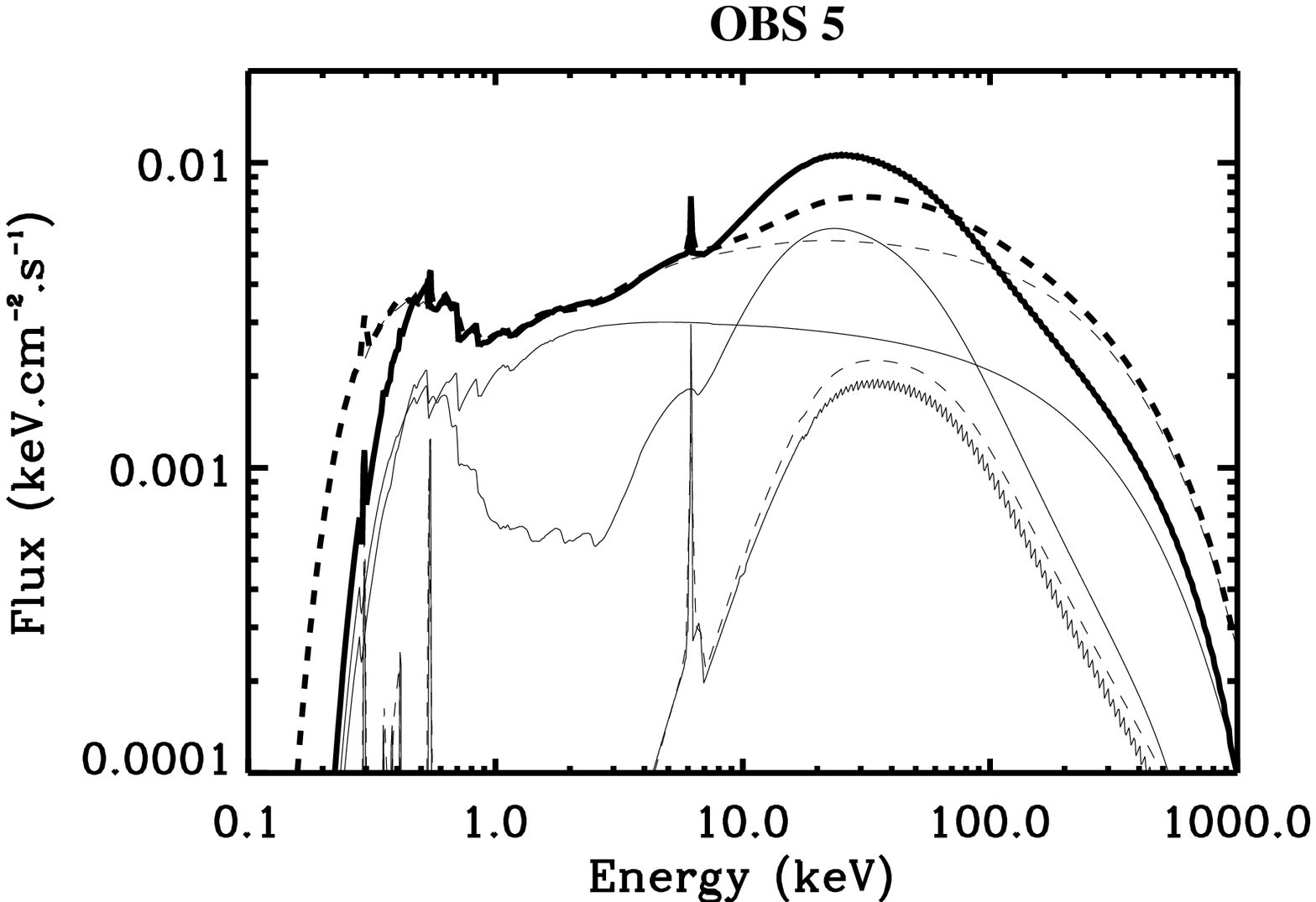}\\
\includegraphics[width=\columnwidth]{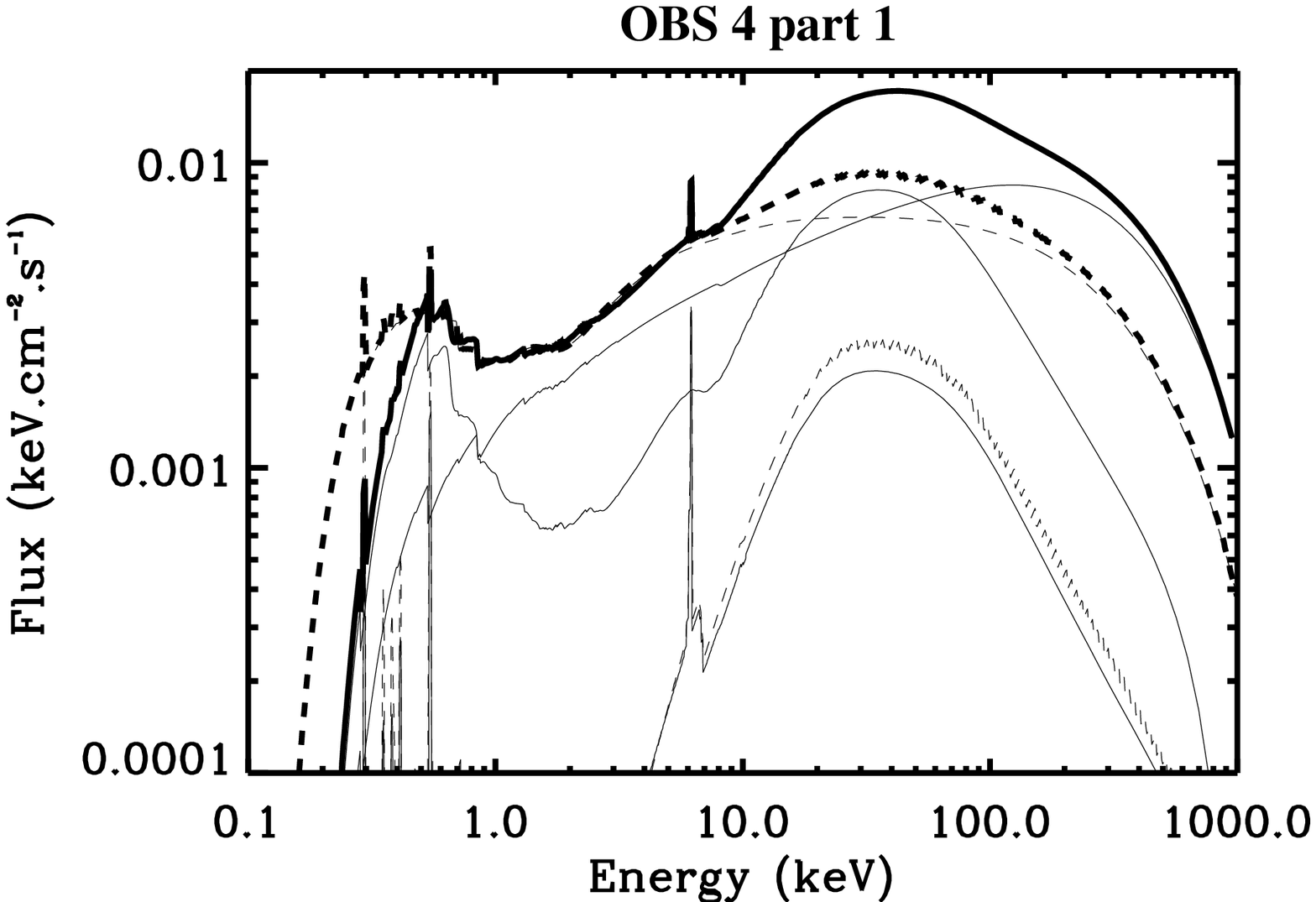}&
\begin{minipage}{0.9\columnwidth}
\vspace*{-6cm}
\caption{ Comparison of the unfolded best fits obtained with the \r and \a models for OBS 1, the first part of OBS 4 and OBS 5. The thick solid lines correspond to the total best fit \r model and the dashed thick line to the \a one. Over-plotted in these figures  in thin lines are the different components of each model.  Thin solid lines: cut-off power law continuum, relativistically blurred ionized reflection and neutral reflection. Thin dashed lines: cut-off power law continuum, modified by relativistically smeared ionized absorption, and neutral reflection. The fluxes are in keV cm$^{-2}$ s$^{-1}$=1.6 $\times$ 10$^{-9}$ erg cm$^{-2}$ s$^{-1}$. \label{eemo}}    
\end{minipage}\\  
\end{tabular}
\end{figure*}

\subsubsection{The best fit results}
\label{bestfitres}
While these models reproduce relatively well the global spectral shape of the data, the fits are not very good. The residuals show some narrow features especially in the soft band and suggest the presence of absorption. These features are even more pronounced in 2005. Such absorbing material is indeed confirmed by a quick look to the RGS data with the presence of absorption features due to the Unresolved Transition Array of Fe VII-XII. 
There is also apparently some variability between OBS 4 and OBS 5 indicating some variations of the absorber properties.  A detailed analysis of this component is out of the scope of the present paper and  will be done in a forthcoming publication (Longinotti et al. in preparation). For the present analysis, we model these absorption features  by adding a warm absorber (WA) component ({\sc absori} in {\sc xspec}) in our fits letting the absorber hydrogen column $N_h^{WA}$ and ionization state $\xi^{WA}$ free to vary. The improvement is highly significant. For \r the $\Delta\chi^2$ is  30, 72 and 75 for the addition of two parameters  for OBS 1/2/3, OBS 4 part 1/2/3 and OBS 5 respectively.   For \a the $\Delta\chi^2$ is equal to 15, 106 and 70. 

The best fits for both models become now statistically acceptable. The corresponding best fit parameter values for the power law continuum, the neutral reflection and the WA  are reported in Tabs \ref{tabcontinuumkdbwa} and \ref{tabcontinuumswindwa} for  \r and \a respectively. The best fit parameter values characterizing the ionized reflecting or absorbing  material are reported in Tabs. \ref{tabkdbwa} and \ref{tabswindwa} . \\

Interestingly, not only the best fits with \r and \a are statistically acceptable, they are also statistically similar and cannot be ruled out with the present XMM-{\it Newton} data. However they predict different spectral characteristics. For instance, the continuum spectral variability between  2001 and 2005 is significantly larger with \r ($\Delta\Gamma$=0.9) compared to \a ($\Delta\Gamma$=0.3). The former case is relatively unusual for a Seyfert 1 galaxy and would require strong changes in the emitting regions. The properties of the WA are also different, at least for OBS 1/2/3 and OBS 4, the column density and ionization parameters being  larger by a factor 10 (and even more for $\xi^{WA}$) between \r and {\sc ABS}.    On the other hand, the neutral reflection normalization is consistent with a constant between the different pointings with both models, supporting the presence of remote reflection. 

For comparison we have plotted the unfolded best fits obtained with \r and \a on Fig. \ref{eemo} for OBS 1, the first part of OBS 4 and OBS 5. We have also over-plotted  the different spectral components like the neutral reflection and, for the \r model, the relativistically blurred ionized reflection. Large differences above 10 keV are expected especially for 2001 and Jan. 2005.\\

We can also analyze more precisely the results obtained for each model separately, beginning with {\sc REF}. Interestingly, while this model requires  still extreme values of the disc inner radius, the disc emissivity power law index, now also constrained by the soft excess, is close to 4 in OBS 4 part 1/2/3 and OBS 5, in better agreement with theoretical expectations (e.g. \citealt{mart00}). This contrasts with the results obtained with the {\sc diskline} model where larger values of $q$ are generally found  (see Tab. \ref{tabfitsimultdiskline}). { We agree that these discrepancies can be partly explained by intrinsic differences between the {\sc Laor} and  {\sc diskline} profiles. Note however } that $q$ is still very large ($>$ 8.6) in OBS 1/2/3 { whatever the relativistic line profile used}. The photon index is also  very steep ($\Gamma\sim$2.45) in these observations. Noticeably, constraining $q$ to be smaller than 5 gives a more reasonable photon index with $\Gamma=$2.26$_{-0.05}^{+0.07}$ while the other parameters do not significantly change. The $\chi^2$ is larger ($\chi^2$/dof=804/787 i.e. $\Delta\chi^2$=20) but still acceptable.  
\begin{figure*}[t!]
\begin{tabular}{cc}
\includegraphics[width=\columnwidth]{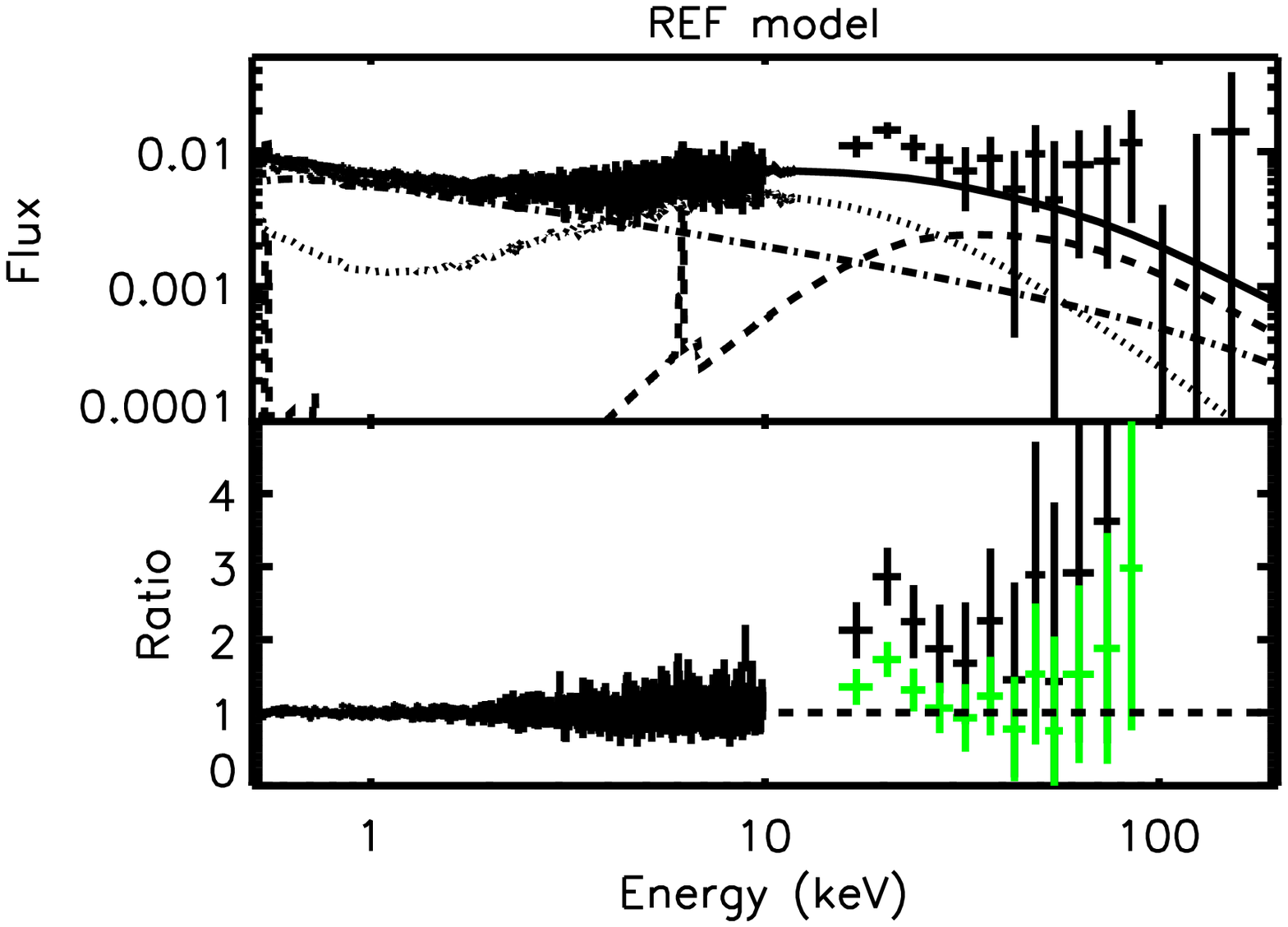}&
\includegraphics[width=\columnwidth]{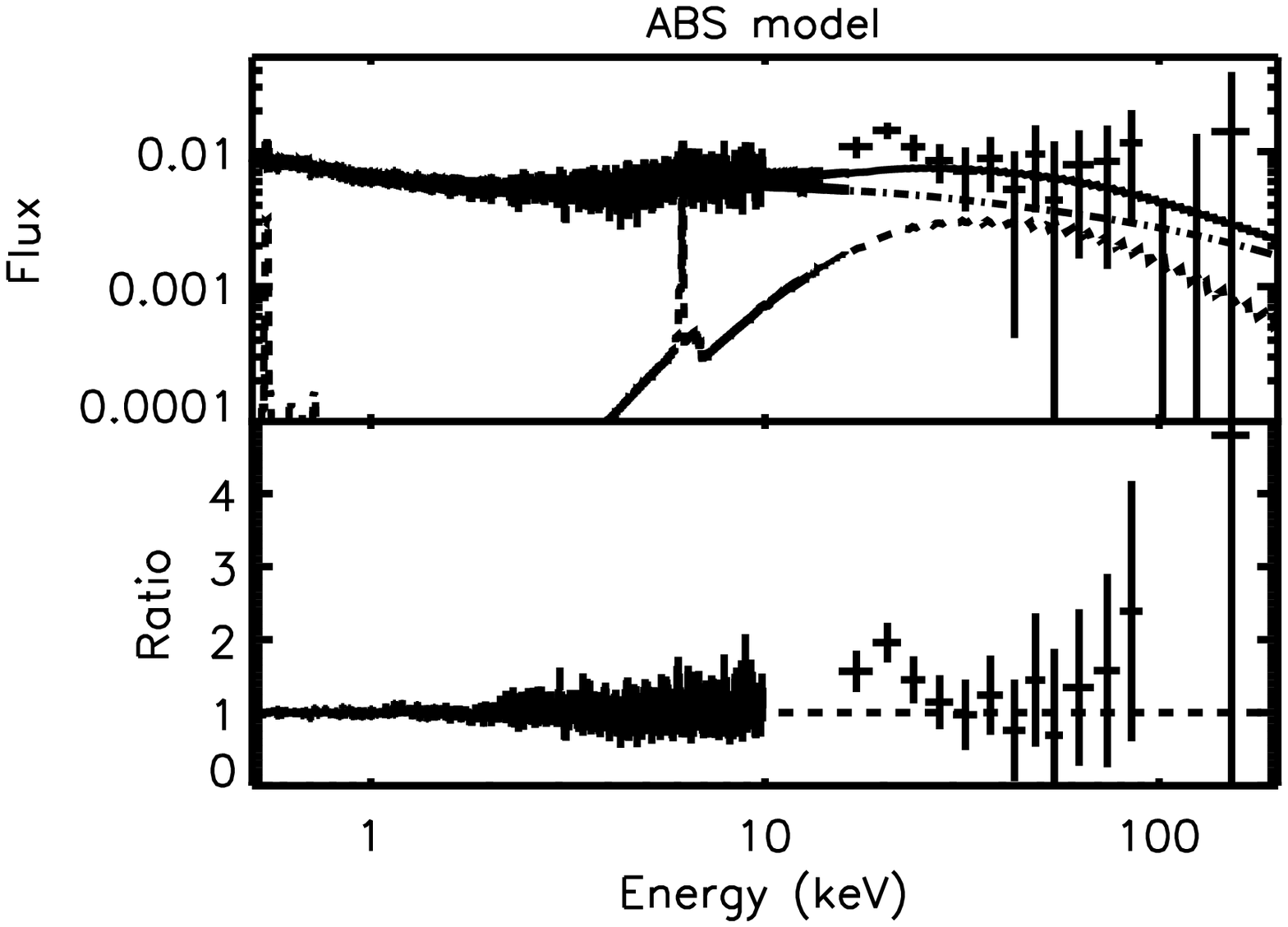}\\
\end{tabular}
\caption{Comparison of the OBS 1 best fit obtained with the \r (left) and \a (right) models  with the simultaneous \BS data. The different spectral components are over-plotted in the upper plots: {\bf left:} neutral reflection (dashed line), ionized blurred reflection (dotted line), cut-off power law (dot-dashed line) and total (solid line). The grey (green on the colored version) ratios correspond to the best fit \r model with the disc emissivity power law index $q$ forced to be smaller than 5 (see Sect. \ref{bestfitres}); {\bf right:} neutral reflection (dashed line), cut-off power law modified by smeared absorption (dot-dashed line) and total (solid line). \label{xmmsax}}
\end{figure*}

We have also reported in Tab. \ref{tabkdbwa} the reflection fraction, i.e. the ratio between the flux in the ionized reflection component and the total flux.
It is of the order of 30-50\%   which is relatively large since we expect $\sim$ 10-20 \% in the case of an isotropic illumination. Finally the properties of the WA significantly change for OBS 5 compared to the other observations.\\ 

On the other hand, the characteristics of the WA are relatively constant between the different pointings with the \a model. The continuum power law index is also close to the standard one for a Seyfert galaxy i.e. $\sim$1.9, reaching also a steeper value ($\sim$ 2.2) in 2001.  To fit the data the \a model requires a change in the properties of the relativistically smeared absorbing material   between 2001 and 2005, the column density and the velocity smearing being larger in 2001. These parameters are more consistent with each other between OBS 4 and OBS 5.\\

\subsubsection{Comparison with a simultaneous \BS observation}

The \BS instruments cover the 0.1-200 keV range and thus better constrains the reflection component compared to XMM-{\it Newton}.  \object{Mkr 841} was pointed by \BS between the 11th and 14th of January 2001, i.e. partly simultaneously with the XMM-{\it Newton} observations of 2001, for a total net time exposure of $\sim$90 ks for the MECS instrument (2-10 keV energy range)  $\sim$40 ks for the PDS (10-200 keV energy range) and only 20 ks for the LECS (0.1-2 keV energy range). Since the \BS observation was much longer than the XMM-{\it Newton} one (see Fig. 1 of \citealt{pet02}), we did not try to fit the XMM-{\it Newton} and \BS data altogether. We only checked the consistency of the \BS data above 10 keV with the XMM-{\it Newton} best fit model expectation.

The unfolded best fit spectra obtained with \r  and \a for OBS 1 are plotted in Fig. \ref{xmmsax} with the \BS MECS and PDS data (we ignore the LECS data due to their poor quality). Only the MECS normalization was let free to vary, the PDS one being fixed to 0.86 the MECS one (following \citealt{fio99}). The agreement is relatively good given the possible variability of the reflection component during the \BS pointing and the fixed value of 300 keV of the power law continuum high energy cut-off in the fits of the XMM-{\it Newton} data. We note however a better agreement of the \a model with the \BS/PDS data above 10 keV. With the \r model, the flux above 10 keV is underestimated by a factor $\sim$2. However this depends  on some of our model assumptions. For example constraining $q$ to be smaller than 5 instead of being completely free (see Sect. \ref{bestfitres}) gives a better agreement (the green data points  on the left plot) with the \BS/PDS data, similar to the agreement found with {\sc ABS}.\\

\section{Indication of redshifted narrow iron lines }
\label{redline}

\subsection{A strongly redshifted line at 4.8 keV in OBS 4}
 
\begin{figure}[!b]
\includegraphics[width=\columnwidth]{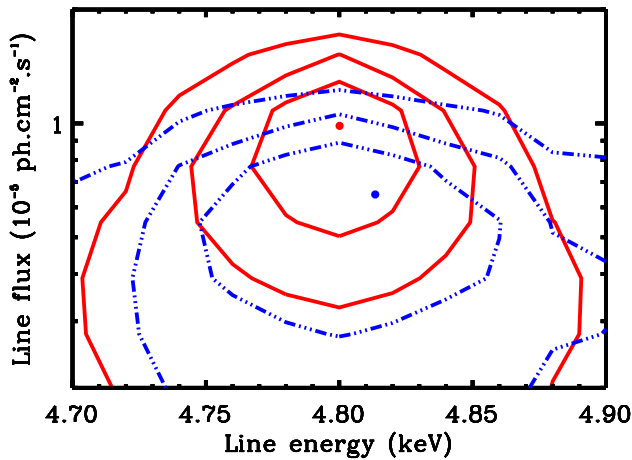}
\caption{Contour plot energy (source frame) vs flux of the line detected close to 4.8
  keV during part 1 of OBS 4. The contours correspond to 68, 90 and 99\% The solid and dash-dotted contours correspond to the EPIC-pn and EPIC-pn + EPIC-mos data respectively}
\label{contplotredline}
\end{figure}
\begin{figure}[!b]
\includegraphics[width=\columnwidth]{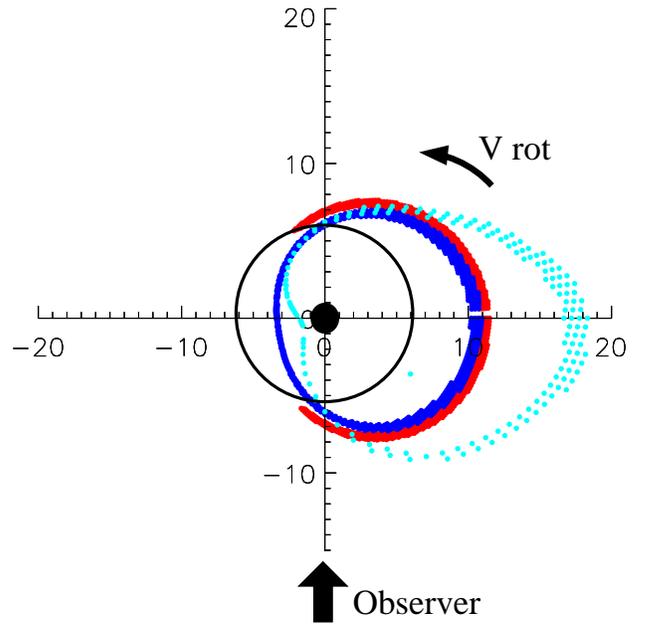}
\caption{A schematic map of the inner accretion disc showing the allowed
  regions where a narrow line with parameters consistent with the one
  observed during part 1 ($E_{line}=4.79^{+0.04}_{-0.02}$ keV) could be
  emitted. In red: $a=$0 and $i=$25. In blue: $a=$0.998 and $i=$25. In
  cyan: $a=0.998$ and $i=$60 deg. The black circle correspond to a
  radius of 6 $r_g$. The disc is rotating counterclockwise.}
\label{mapdisc}
\end{figure}
We focus here on a possible detection of a narrow
line at 4.8 keV in the first part of OBS 4. This feature is clearly
visible on Fig. \ref{ratioobs4po} but also on the excess map shown on Fig. \ref{mapexcess}.
To estimate its confidence level we first added a gaussian with energy
$\sim$ 4.8 keV to a simple power law best fit model of OBS 4 part 1. The gaussian energy and flux are let free to
vary but, given the apparent narrowness of the feature, we fix the
gaussian width to zero. The excess is clearly detected with an
improvement of the fit of $\Delta \chi^2= 11$. The best fit parameters of the line are
$E_{line}=4.80^{+0.03}_{-0.03}$ keV (source frame) and an equivalent width $EW=50\pm20$
eV. The contour plot of the line energy versus the line flux is plotted in solid line on
Fig. \ref{contplotredline}. 

We also estimate the confidence level of this detection by simulating a
large number of spectra with a continuum similar to that of OBS 4 part 1. We
fit the different simulations with a simple power law. We then add a
gaussian, fixing its energy between 4 and 8 keV by step of 0.1 keV and
fit again the data, the gaussian normalization being free to vary while
the gaussian width was fixed to 0. Then for each simulated spectra, we
keep the best $\Delta\chi^2$ resulting from this procedure. For 1000
simulations we only find 14 cases
where a $\Delta\chi^2 >$11 corresponding to a confidence level of 98.5\%.
The strong detection in the EPIC-pn data is however attenuated by the weak
detection in the MOS data. Indeed we have plotted  in dashed line on
Fig. \ref{contplot} the corresponding contour plot when fitting the EPIC-pn
and MOS data simultaneously. The detection is now significant at only 84.4\% from our simulations.\\

If we assume that this line is a redshifted neutral fluorescent iron
line, its observation at 4.8 keV (source frame) implies a redshift factor
of $\sim$ 0.75. Such high redshift value suggests an origin close to the
central engine where relativistic effects become important. Moreover,
the narrow shape of the line constrains the emitting region to be
sufficiently small. There are growing
evidences of such variable narrow emission features in the X-ray spectra of
several AGN in the literature \citep{tur04,por04,pet02,tomb07}.  The common scheme to explain these components
suppose the presence of transient magnetic flares briefly illuminating a
localized part (hot spot) of the accretion disc and producing the iron line by
fluorescence. 

Following Longinotti et al. (2004, Fig. 4) we have produced
a 90\% probability map (from the 90\% error on the line energy) of the accretion disc where the line could come
from, for different values of the black hole specific momentum and disc
inclination angles (cf. Fig. \ref{mapdisc}). 
For an inclination of 25 deg. the line emission can come from region at a distance as small as 3 $r_g$ from the central black hole and even closer for larger inclinations. Moreover, if we assume that we observed the same hot spot on the disc for more than an orbit then the global line profile is expected to be relatively complex (e.g. \citealt{dov04b}) with a strong blue peak. If we  interpret the observed feature with this blue peak, a Kerr solution is unavoidable \citep{pec05}.

\subsection{A slightly redshifted line at 6.2 keV in OBS 1}
\begin{figure}[!t]
\includegraphics[width=\columnwidth]{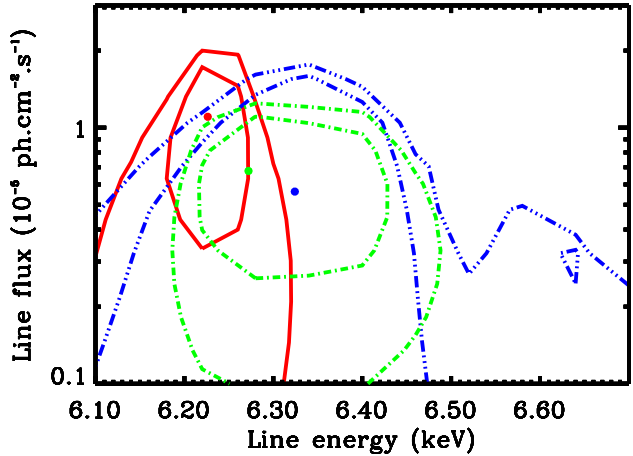}
\caption{Contour plot (68 and 90 \%) energy (source frame) vs flux of the narrow line ($\sigma$=0 eV) added to the {\sc diskline} fit of OBS 1 (solid red line), OBS 2 (dotdotdot-dashed blue line) and OBS 3 (dot-dashed green line). The narrow line energy in OBS 1 peaked close to 6.2 keV (source frame)}
\label{contplotredline2}
\end{figure}

We discuss now the possible detection of a narrow line feature near 6.2 keV (source frame) in the first observation of 2001 as already noted in Sect. \ref{remoteref}. Starting from the best fit {\sc diskline} model reported in Tab. \ref{tabfitsimultdiskline}, we add a narrow ($\sigma$=0 eV) gaussian line letting its energy free to vary for each data set OBS 1, OBS 2 and OBS 3. This differs from the analysis done  in Sect. \ref{remoteref} where the narrow gaussian line parameters were kept constant between the different observations. The corresponding energy vs. flux contour plot of each narrow line are reported in Fig. \ref{contplotredline2}.
The narrow line energy is consistent with the neutral fluorescent iron line energy of 6.4 keV in OBS 2 and 3 but is inconsistent at more than 90\% with this value for OBS 1 where $E_{line}=6.23_{-0.03}^{+0.05}$ keV.

\begin{figure}[!t]
\includegraphics[width=\columnwidth]{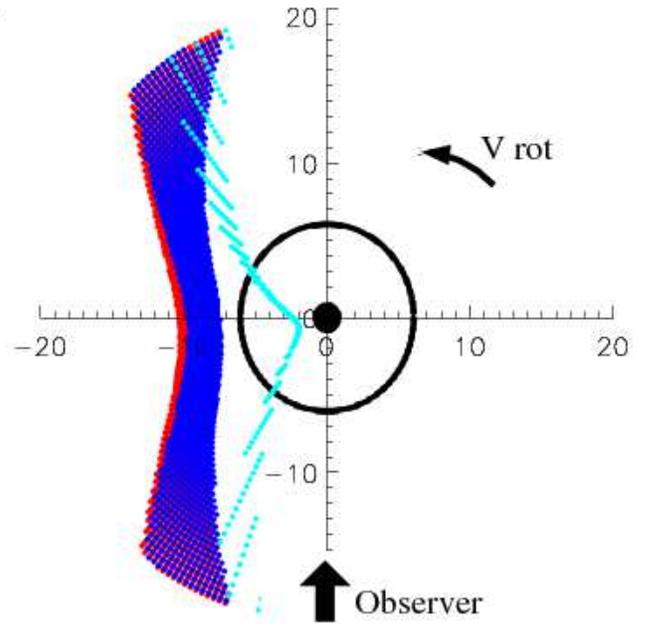}
\caption{Same as Fig. \ref{mapdisc}  but for a narrow line with parameters consistent with the one
  observed during OBS 1 ($E_{line}=6.23_{-0.03}^{+0.05}$ keV). We have limited ourselves to the inner part ($<$ 20 $r_g$) of the accretion disc. Cf. Fig. \ref{mapdisc} for the color scheme. The blue and red regions overlapped almost totally.}
\label{mapdiscb}
\end{figure}

If interpreted, as above, as a (slightly) redshifted narrow iron line we can also produced
a 90\% probability maps of the accretion disc where the line could come
from, for different values of the black hole specific momentum and disc
inclination angles. This has been done in Fig. \ref{mapdiscb}. 

\section{Summary and Discussion}
\label{disc}

We have presented in this paper a detailed spectral analysis of the XMM-{\it Newton}/EPIC-pn data of  \object{Mkr 841} including all the XMM-{\it Newton} observations of this object. Given the relative spectral complexity of this source we would like to summarize first the different results we obtained.

 \object{Mkr 841} has been observed three times in January 2001 (OBS 1, OBS 2 and OBS 3) and two times in 2005 in January (OBS 4) and July (OBS 5). Flux and spectral variations being present during the $\sim$ 45 ks of OBS 4, this observation has been divided in 3 parts for the spectral analysis. A strong soft excess as well as a complex iron line profile, both known to be present in this source for a long time, are clearly detected in all these pointings, the high sensitivity of the XMM-{\it Newton}/EPIC-pn  unveiling very puzzling spectral and temporal behavior. The 0.5-10 keV flux varies by a factor 3 in 4 years and is dominated by the soft band ($<$ 3 keV) variability, the data above $\sim$5 keV keeping roughly constant  on short and long time scales. The spectral variability is also important. Fitting the 3-10 keV data with a simple power law the spectral index varies from $\sim$1.9 in 2001 down to $\sim$1.3 in Jan. 2005. However a simple pivoting power law component cannot explain by itself the broadband 0.5-10 keV spectral variability observed on year time scale thus indicating a more complex spectral variability. 

The line profile is also complex, being apparently a mixture of broad and narrow components. It is highly variable and the 2005 data confirm the rapid line variability observed in 2001 by \citet{pet02} and \citet{lon04} with a variability time scale as short as a few kilo-seconds. This strong X-ray variability suggests a small emitting region  close to the black hole. Fitting the line with a {\sc diskline} model gives a good representation of the line profile and requires in some cases steep disc emissivity power law indexes. 

Given the high complexity of the spectra, we choose to decrease the degree of freedom of some of our fits by fitting simultaneously observations with roughly the same underlying continuum, e.g. the three pointings of 2001 as well as the different parts of OBS 4. In these cases,  during the fits the power law continuum was kept constant in shape but not in flux between the data sets.  Concerning OBS 5, it was analyzed normally as a single observation. A narrow line component is present in all the pointings and its flux is consistent with a constant between 2001 and 2005. This suggests the presence of remote reflection in the data. If this interpretation is correct, then  the observed variability of the iron line complex is more likely due to the variability of the underlying component below the narrow iron line, either due to a broad line component or strong smeared absorption.

Then we analyze the broad band (0.5-10 keV) EPIC-pn energy range. We applied two different models, recently proposed in the literature, for the origin of the soft excess: a relativistically blurred ionized reflection ({\sc REF}) and a relativistically smeared ionized absorption ({\sc ABS})). We also added a neutral reflection to reproduce the narrow line component. Both models are a reasonable representation of the
overall broad band shape but are formally statistically unacceptable
du to the presence of absorption features in the soft energy range. The addition of a warm absorber strongly improves the fits that converge to statistically acceptable and statistically equivalent representations of the data. Both models are also consistent with the partly simultaneous \BS observations done in 2001.

Finally we also note the presence of a strong narrow feature near 4.8 keV in the first part of OBS 4. Its detection is marginally significant in the EPIC-pn data ($>$98.5\%) but the significance decreases to 84\% when we include the MOS data. If interpreted as the blue horn of a relativistically distorted neutral iron line, the large redshift implies the presence of a Kerr black hole.\\

\begin{figure}[t!]
\includegraphics[width=\columnwidth]{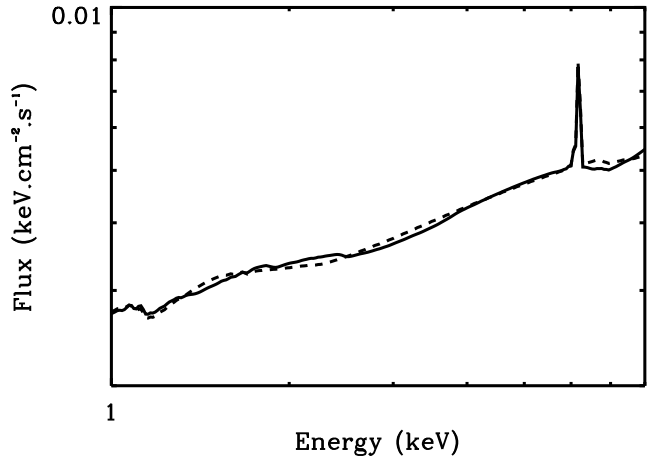}
\caption{Unfolded best fits of OBS 5 obtained with \r (solid line) and \a (dashed line) in the 1-8 keV range. The curvature of the spectral shape below to the iron line, due to a broad line component in \r and  by smeared absorption in  {\sc ABS}, is present in both cases.\label{zoom}}
\end{figure}

The most remarkable result of our study is  to reproduce reasonably well the broad band (0.5-10 keV) and complex spectral characteristics of  \object{Mkr 841} with a small number of spectral components. However we were not able to discriminate between the two different interpretations of the soft X-ray excess, which have been debated in the recent literature. Moreover both models are able to reproduce the spectral shape close to the iron line either by adding a broad component, like in the case of the \r model, or by modifying the power law shape, like in the the \a model. This is exemplified in Fig. \ref{zoom} where we show a zoom of the unfolded best fit models obtained with \r and \a for OBS 5 in the 1-8 keV range. 

Whatever the model used, the need of strong relativistic effects to smear/blur the absorption/emission features suggest an origin of the X-ray emission close to the central black hole. Then the absorbing material required in the \a model could be the base of a disc wind/jet surrounding the inner X-ray emitting region. The external part of this wind, with lower velocities, may then explain the absorption features of the WA.

On the other hand, the \r model is based on the presence of strong reflection as indicated by the large reflection fraction that we obtained to fit the data. Another way of quantifying this large amount of reflection is to measure the commonly used reflection parameter $R$.
Since this is not a direct parameter of the reflection tables that we used, we estimate its value by roughly adjusting  the reflection shape (ignoring the blurring effects) with the one expected with the {\sc pexriv} model of {\sc xspec} fixing the power law photon index and normalization as well as the ionization parameter of  {\sc pexriv}  to our best fit values. Following this method, we obtain $R$ values of the order of  10,  2 and  4 for OBS 1/2/3, OBS 4 and OBS 5 respectively. We recall here that the blurring effects, as well as the ionization level of the reflecting material considerably smooth the reflection shape.  In consequence even a reflection component with $R=10$ is strongly attenuated in the outgoing spectrum. 
But such large reflections necessarily require the presence of some effects  that increase the reflection component compared to the illuminating continuum like e.g. in the light bending effects  model \citep{min03,min04b} or the inhomogeneous accretion flows model proposed recently by \citet{mer06}.\\

{ The low statistics of our RMS spectra (cf. Fig. \ref{rms}) do not allow a detailed comparison with
the variability behavior expected with \a or \r and thus unable us to discriminate the two models. We can note however that, while both models can correctly reproduce  the RMS spectra of some AGNs (e.g. \citealt{gie06,pon06}),  \a predicts an RMS spectrum which rapidly increases and  peaks around 1 keV, where the absorber enhances the variability \citep{gie06}. This is not seen in our RMS spectra  and so argues against this model a bit or at least that there is no absorber variability on short timescales. But this has to be tested on data of better quality.}


Finally, we note that our (admittedly marginal) detection of redshifted narrow iron lines in the data better fits in the relativistically blurred reflection interpretation where such features are naturally expected. But we agree that this is a relatively weak argument that requires further investigations.

\section{Conclusion}

 \object{Mkr 841} is a bright Seyfert 1 galaxy which is known to possess a strong soft excess and complex iron line profile. We have presented a detailed analysis of the whole  XMM-{\it Newton}/EPIC-pn data of this object from 2001 to 2005. The conclusions of this analysis can be summed up as follows:
\begin{itemize}
\item  Strong flux and spectral variability are observed on month and year time scales. The 0.5-10 keV flux varies by a factor 3 in 4 years and is dominated by the soft band ($<$ 3 keV) variability, the data above $\sim$ 5 keV keeping roughly constant  on short and long time scales. 
\item The iron line is apparently a mixture of broad and narrow components. Its profile is rapidly varying on very short time scale (a few ks). 
\item The broad band 0.5-10 keV spectrum is well described by a model including 1) a neutral absorption,  2) a cut-off power law continuum,  3) a
neutral reflection and a fourth component for the soft excess. We use two different models for this last component:  a relativistically blurred photoionized
  reflection (\r model) and a relativistically smeared ionized absorption (\a model). Both models give statistically acceptable and statistically equivalent fits of the data even including, for 2001 observations, partly simultaneous \BS data up to 200 keV.
\item Both models agree with the presence of remote reflection characterized by a constant narrow component in the data. However they differ on the presence of a broad line component  present in \r but not needed in  {\sc ABS}. Consequently, the physical interpretation of the line profile variability is quite different, resulting from the variability of the broad line component in \r and from the variability of the absorbing medium in  {\sc ABS}).
\item  \object{Mkr 841} is also a good candidate for the observation of redshifted narrow iron lines, two marginal detections being discussed in this paper. If such features would favor the relativistically blurred photoionized  reflection interpretation, their marginal detections do not permit any clear conclusion.
\end{itemize}  

While it seems reasonable that  the reality is a complex combination of absorption and reflection effects the present analysis underlines the difficulty in well disentangling these components in the X-ray spectra of AGN. Broad band observations including data above 10 keV (Suzaku, INTEGRAL) will certainly help to better test both interpretations but the high sensitivity in the 10-50 keV energy range is crucial. We may need to wait for missions like SIMBOL-X or XEUS to have a clear  understanding of what processes really take place in sources like  \object{Mkr 841}.

\end{document}